\documentclass{aa}

\usepackage{graphicx}

\newcommand{\bv}{(\textrm{B$-$V})}
\newcommand{\vs}{(\textrm{V$-$S})}
\newcommand{\yv}{(\textrm{Y$-$V})}
\newcommand{\yz}{(\textrm{Y$-$Z})}
\newcommand{\zv}{(\textrm{Z$-$V})}
\newcommand{\bbv}{(\textrm{B$_{2}-$V$_{1}$})}
\newcommand{\bbg}{(\textrm{B$_{2}-$G})}
\newcommand{\bbb}{(\textrm{B$_{1}-$B$_{2}$})}
\newcommand{\ttt}{\textrm{t}}
\newcommand{\vr}{(\textrm{V$-$R})_{(\textrm{\tiny C})}}
\newcommand{\ri}{(\textrm{R$-$I})_{(\textrm{\tiny C})}}
\newcommand{\vi}{(\textrm{V$-$I})_{(\textrm{\tiny C})}}
\newcommand{\ddoa}{\textrm{C(42--45)}}
\newcommand{\ddob}{\textrm{C(42--48)}}
\newcommand{\ddoc}{\textrm{C(45--48)}}

\newcommand{\feh}{\textrm{[Fe/H]}}

\begin{document}

\title{
IRFM temperature calibrations for the Vilnius, \\
Geneva, RI$_{(\textrm{\small C})}$ and DDO photometric systems
\thanks{Based on data from the GCPD.}
}

\author{
Jorge Mel\'{e}ndez \inst{1,2} \and Iv\'{a}n Ram\'{\i}rez \inst{1}
}

\institute{
Seminario Permanente de Astronom\'{\i}a y Ciencias
Espaciales, Universidad Nacional Mayor de San Marcos. \\
Ciudad Universitaria, Facultad de Ciencias F\'{\i}sicas. Av.
Venezuela s/n, Lima 1--Per\'{u}. \\
\email{jorge@astro.iag.usp.br; ivan@fastmail.ca} \and Instituto de
Matem\'{a}tica y Ciencias Afines, Jr. Ancash 536, Lima
1--Per\'{u}.}

\offprints{J. Mel\'{e}ndez.}

\date{Received 02 October 2002 / Accepted 07 November 2002}

\abstract{We have used the infrared flux method (IRFM)
temperatures of a large sample of late type dwarfs given by Alonso
et al. (\cite{alonso:irfm}) to calibrate empirically the relations
$T_{eff}=f$(colour, [Fe/H]) for the Vilnius, Geneva,
RI$_{(\textrm{\tiny C})}$ (Cousins) and DDO photometric systems.
The resulting  temperature scale and intrinsic colour-colour
diagrams for these systems are also obtained. From this scale, the
solar colours are derived and compared with those of the solar
twin 18 Sco. Since our work is based on the same $T_{eff}$ and
[Fe/H] values used by Alonso et al. (\cite{alonso:escala}) to
calibrate other colours, we now have an homogeneous calibration
for a large set of photometric systems. }

\authorrunning{J. Mel\'{e}ndez  \& I. Ram\'{\i}rez}
\titlerunning{IRFM temperature calibrations}

\maketitle

\section{Introduction}

Photometric calibrations of effective temperature can be regarded
as fundamental tools to compare observations with theoretical
models. They are needed to transform theoretical HR diagrams into
colour-magnitude diagrams (e.g. VandenBerg \cite{vdb}) as well as
to perform the colour synthesis of stellar populations (e.g.
Bruzual \& Charlot \cite{bruzual}). Chemical abundance studies of
stars, on the other hand, use the effective temperature
($T_{eff}$) as a key parameter, so that reliable methods to
calculate them are widely required (e.g Mel\'{e}ndez \& Barbuy
\cite{mb2002}; Nissen et al. \cite{nissen}; Ram\'{\i}rez \& Cohen
\cite{rc}).

Since the first published photometric calibration of $T_{eff}$
(Popper \cite{popper}) many other papers devoted to the study of
the colour relations between $T_{eff}$, [Fe/H] and even $\log g$
have been written (e.g. Code et al. \cite{code}, Bell \&
Gustaffson \cite{bell89},  Bessell et al. \cite{bessell},
Houdashelt et al. \cite{houdash}). Most of them, unfortunately,
considered only a relatively low number of stars, the problem
being more critical at the metal--poor end. To make this problem
even worse we shall mention that these different calibrations used
different sets of $T_{eff}$ and [Fe/H] values, thus yielding
inhomogeneous empirical temperature scales.

Alonso et al. (\cite{alonso:escala}; hereafter AAM96b) solved the
problem partially by performing calibrations for the
BVRI$_{(\textrm{\tiny J})}$, JHK$_{(\textrm{\tiny TCS})}$ and
uvby-$\beta$ colours with a large sample of cool dwarfs covering
the ranges $-3.0<$[Fe/H]$<+0.5$ and 3500 K$<T_{eff}<8000$ K. The
effective temperatures for the stars of their sample were computed
by means of the infrared flux method (IRFM) whilst the
corresponding metallicities were taken from the Cayrel de Strobel
et al. (\cite{cayrel92}) catalog and partially from previously
published photometric calibrations as described in Alonso et al.
(\cite{alonso:irfm}; hereafter AAM96a).

Even though it is possible to obtain the effective temperature of
a star by applying the calibrations mentioned above, in some cases
the only colour available is $\bv$, which is very sensitive to
metallicity, or colours that belong to photometric systems that
lack a homogeneous calibration. In this sense, the aim of this
work is to extend the calibrations given in AAM96b to other
colours for which a large amount of photometric data exist, like
the Geneva system.

In this paper, we present the calibrations for the Vilnius,
Geneva, RI$_{(\textrm{\tiny C})}$ and DDO photometric systems
obtained with the sample in AAM96a and compare them to previously
published calibrations. In Sect. 2 the sample and adopted
photometry are described. The calibrations and the comparison with
other works are presented in Sect. 3 leaving the discussion of the
resulting temperature scale to Sect. 4. Conclusions are summarized
in Sect.~5.

\section{The sample}

Although the atmospheric parameters $T_{eff}$ and $\feh$ of the
sample used in this work are well described in AAM96a, we shall
mention some extra details on these.

All the stars of the sample are supposed to be late type dwarfs
and subdwarfs. Nevertheless, a revision of their $\log g$ values
in the Cayrel de Strobel et al. (\cite{cayrel2001}) catalog and
their spectral types in SIMBAD database enabled us to identify
some of these stars as giants and others as early type stars. We
also found some variable and double stars.

Several stars, apart from those belonging to the groups listed
above, were excluded from our calibrations due to anomalies in
their colours: G245-032, G009-016, G043-003, G014-039, G141-019,
HD181007, HR8832, G190-015, G217-008 (anomalous Vilnius colours);
HR66, G026-012 (anomalous Geneva colours); vA 560 (anomalous
Cousins colours); HD45282, HR4623, G154-021 (anomalous DDO
colours). Stars with kinematical assignment of metallicity were
also excluded as in AAM96b.

The photometry adopted in this work was obtained from the General
Catalogue of Photometric Data (GCPD, Mermilliod et al.
\cite{gcpd}). For the sake of completeness we compiled all of the
colours available for each star but, obviously, only a subset of
these proved to be sensitive to the effective temperature, while
being nearly independent of metallicity. Colours satisfying this
condition are: $\vs$ and $\yv$ in the Vilnius system; $\bbv$ and
$\bbg$ in the Geneva system; $\vr$, $\ri$ and $\vi$ in the
RI$_{(\textrm{\tiny C})}$ system and $\ddoa$ and $\ddob$ in the
DDO system. The reddening corrections $E(\textrm{colour})$ were
computed from the $E\bv$ values given in AAM96a and the reddening
ratios listed in Table~\ref{red-ratios}. Unreddened $\yv=\yz+\zv$
colours were obtained from intrinsic $\yz$ and $\zv$ colours. A
similar procedure was employed to obtain unreddened $\ddob$
colours.

\begin{table}
\centering
\begin{tabular}{l c c}
Colour & $E(\textrm{colour})/E\bv$ & Reference \\ \hline
$\vs$ & 0.70 & Strai\v{z}ys (\cite{straizys}) \\
$\yz$ & 0.52 & Strai\v{z}ys (\cite{straizys}) \\
$\zv$ & 0.29 & Strai\v{z}ys (\cite{straizys}) \\
$\bbv$ & 0.75 & Cramer (\cite{cramer}) \\
$\bbg$ & 1.01 & Cramer (\cite{cramer}) \\
$\vr$ & 0.56 & Savage \& Mathis (\cite{sym}) \\
$\ri$ & 0.69 & Savage \& Mathis (\cite{sym}) \\
$\vi$ & 1.25 & Savage \& Mathis (\cite{sym}) \\
$\ddoa$ & 0.23 &  McClure (\cite{mcclure}) \\
$\ddoc$ & 0.31 &  McClure (\cite{mcclure}) \\  \hline
\end{tabular}
\caption{Reddening ratios for late type dwarfs.}\label{red-ratios}
\end{table}

\section{The calibrations}

Following the methodology of AAM96b, we have used the parameter
$\theta_{eff}=5040/T_{eff}$ instead of $T_{eff}$ to perform the
calibrations. In general, the multi-parametric relation
$\theta_{eff}=f(\textrm{colour},\feh)$ is a second order
polynomial. Nevertheless, in order to improve our results, we have
only considered those coefficients that are larger than $3\sigma$.

The residuals of the fits as a function of colour index and
metallicity are often shown to check if systematic errors are
introduced by the calibration formulae and also to illustrate
their ranges of applicability in colour and metallicity.

\subsection{Vilnius colours}

The Vilnius photometric system was originally designed to
facilitate a bi-dimensional spectral classification for stars of
any spectral type, luminosity class and reddening; the selection
of medium-band filters being done in order to observe faint stars.

For the colour index $\vs$, which seems to be the best $T_{eff}$
indicator in the Vilnius system, we found, using data for 120
stars, the following fit:
\begin{eqnarray}
\theta_{eff}=0.248+1.336\vs-0.354\vs^{2} \nonumber
\\ -0.029\feh-0.008\feh^{2}\ ,\label{cali-vs}
\end{eqnarray}
for which the standard deviation $\sigma(\theta_{eff})=0.020$. In
terms of $T_{eff}$ this amounts to 104 K.

The application ranges of Eq. (\ref{cali-vs}) are:
\[0.40<\vs<1.20\ \ \ \textrm{for }\ -0.5<\feh<+0.5\ ,\]
\[0.40<\vs<0.70\ \ \ \textrm{for }\ -1.5<\feh<-0.5\ ,\]
\[0.45<\vs<0.70\ \ \ \textrm{for }\ -2.5<\feh<-1.5\ .\]

The stars used to obtain Eq. (\ref{cali-vs}) whose temperatures
calculated from the calibration formula differ by more than
$2\sigma$ from their IRFM temperatures are: G273-001, G021-006,
G084-029 and HR6136.

\begin{figure*}\centering
\includegraphics[bb=18 16 565 320,width=17cm]{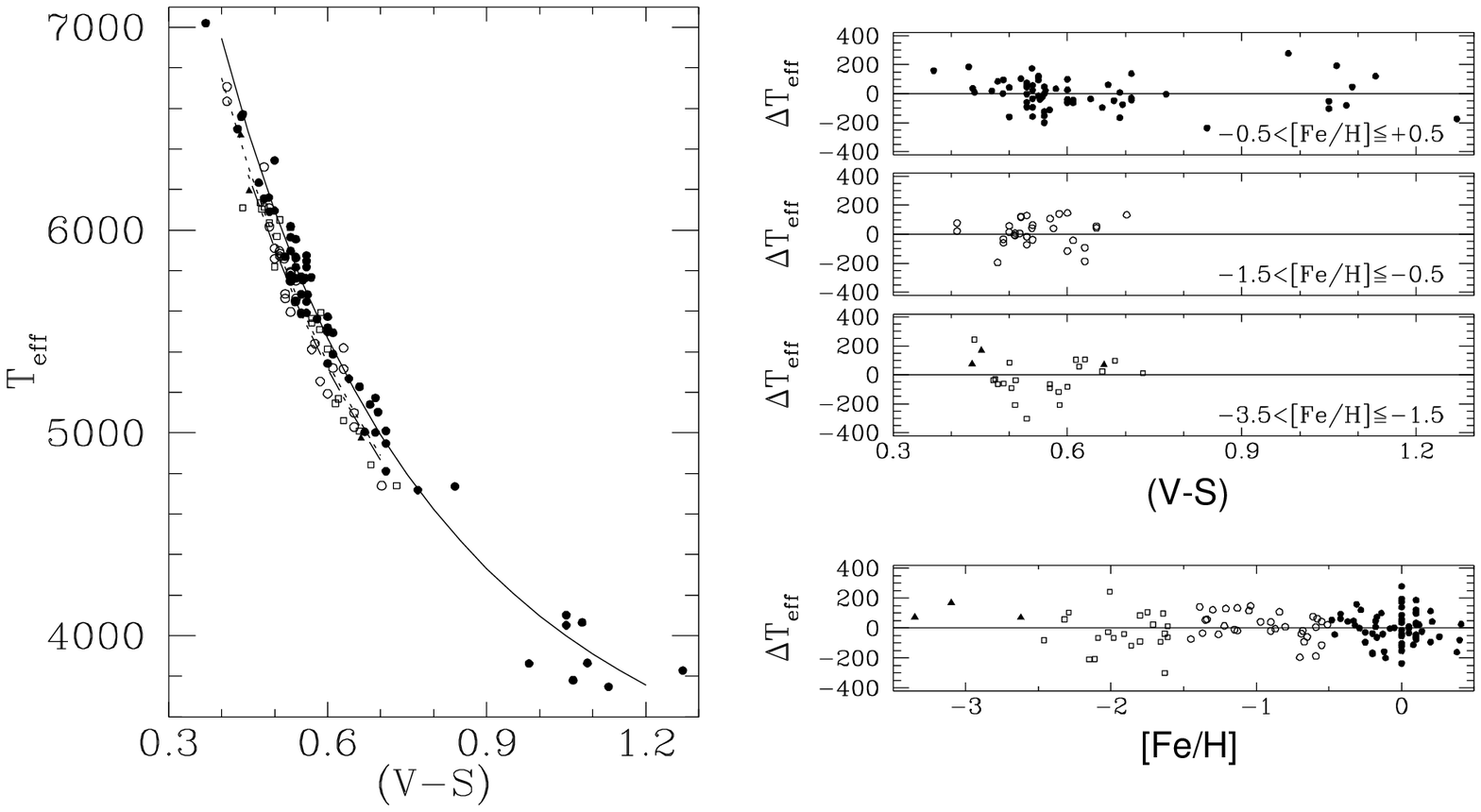}
\caption{Left: $T_{eff}$ vs $\vs$ observed for the metallicity
ranges $-0.5<\feh\leq+0.5$ (filled circles), $-1.5<\feh\leq-0.5$
(open circles), $-2.5<\feh\leq-1.5$ (squares) and
$-3.5<\feh\leq-2.5$ (triangles). The curves corresponding to our
calibration for $\feh=0$ (solid line), $\feh=-1$ (dotted line) and
$\feh=-2$ (dashed line) are also shown. Right: residuals of the
fit ($\Delta T_{eff}=T_{eff}^{\textrm{\tiny
cal}}-T_{eff}^{\textrm{\tiny IRFM}}$) as a function of $\vs$ (for
the metallicity ranges indicated in the lower right section of the
three upper panels) and $\feh$ (bottom panel).} \label{fig:vilvs}
\end{figure*}

Figure \ref{fig:vilvs} shows the sample and the residuals of this
fit as a function of $\vs$ and $\feh$. There is a slight tendency
towards lower temperatures for stars belonging to the $\feh\sim-2$
group but no other trends are apparent.

\begin{figure*}\centering
\includegraphics[bb=18 16 565 320,width=17cm]{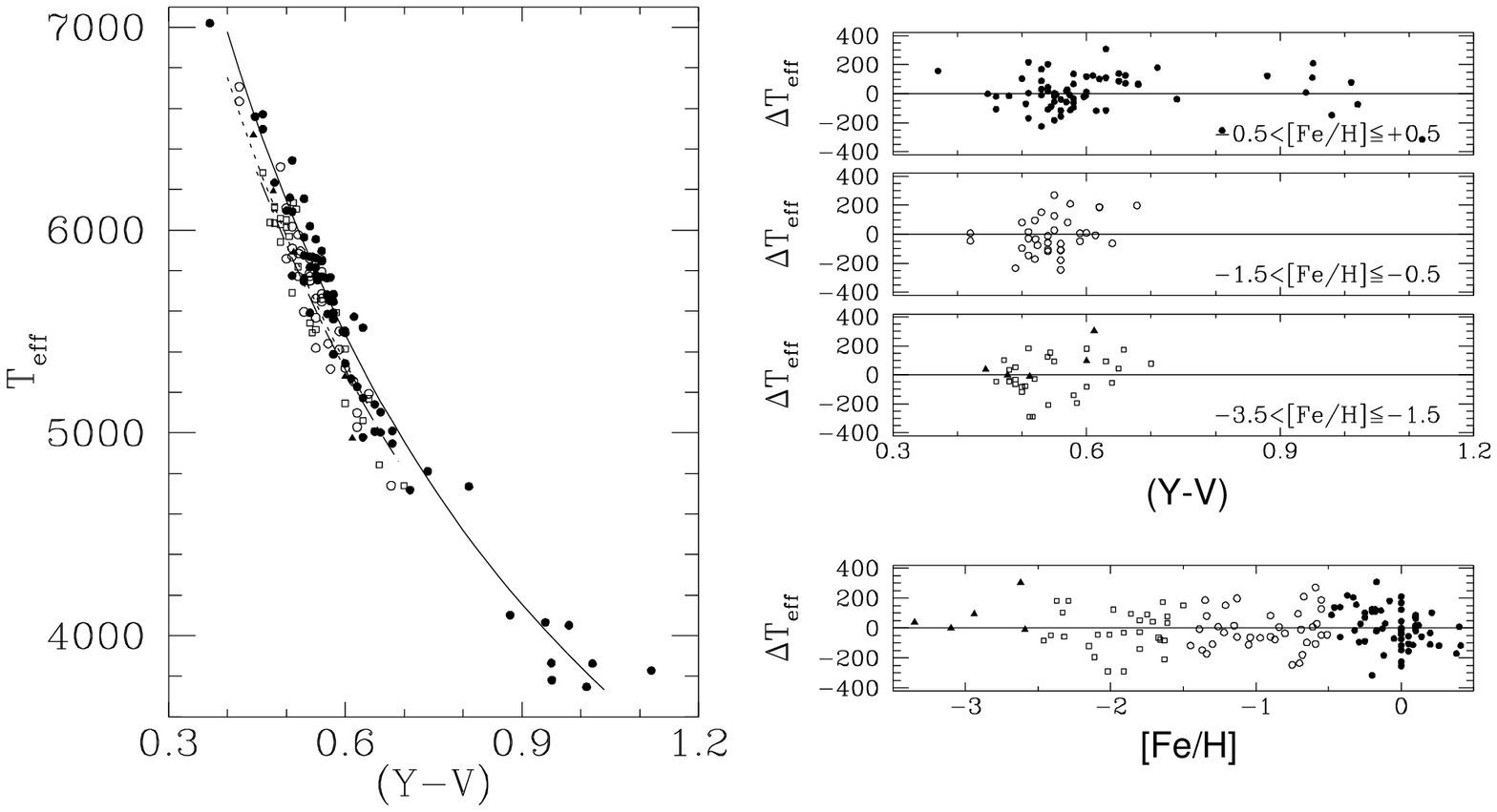}
\caption{The same as in Fig. \ref{fig:vilvs} for $T_{eff}$ versus
$\yv$.} \label{fig:vilyv}
\end{figure*}

The $\yv$ colours, temperatures and metallicities of 134 stars
allowed us to obtain the calibration formula
\begin{eqnarray}
\theta_{eff}=0.330+0.981\yv-0.033\feh \nonumber \\ -0.009\feh^{2}\
,\label{cali-yv}
\end{eqnarray}
with $\sigma(\theta_{eff})=0.024$ (126 K). This formula is
applicable in the ranges:
\[0.40<\yv<1.05\ \ \ \textrm{for }\ -0.5<\feh<+0.5\ ,\]
\[0.40<\yv<0.65\ \ \ \textrm{for }\ -1.5<\feh<-0.5\ ,\]
\[0.45<\yv<0.70\ \ \ \textrm{for }\ -2.5<\feh<-1.5\ .\]

The following stars depart more than $2\sigma$ from the mean fit:
G119-052, G217-008, G029-023, G021-006, G019-027, G014-024 and
HR6752.

Three stars with $\feh<-2.5$ were included in the sample used to
calibrate this photometric system but since they showed a
systematic tendency towards higher temperatures, we decided not to
consider Eqs. (\ref{cali-vs}) and (\ref{cali-yv}) valid for the
metal-poor end.

The high values of the residuals of these fits for the cooler
stars ($T_{eff}\sim4000$ K) as shown by Figs. \ref{fig:vilvs} and
\ref{fig:vilyv} could be associated mainly to rough estimates of
the metallicity of this group.

The gradients $\Delta T_{eff}/\Delta\vs$ and $\Delta
T_{eff}/\Delta\yv$ corresponding to our calibrations are both
nearly independent of metallicity and vary from 70 K per 0.01 mag
for $T_{eff}\sim6000$ K [$\vs\sim0.51$, $\yv\sim0.52$] to 25 K per
0.01 mag for $T_{eff}\sim4000$ K [$\vs\sim1.05$, $\yv\sim0.95$].
As a consequence, an observational error of 0.01 mag in the
photometry implies an error of about 1.2\% for the hottest stars
and 0.6\% for the cool end when using formulae (\ref{cali-vs}) and
(\ref{cali-yv}).

On the other hand, the variations $\Delta T_{eff}/\Delta\feh$
slightly depend on colour for both cases but the dependence of
their values for a given $T_{eff}$ with $\feh$ is similar (see
Table \ref{gradients-feh-vilnius}). For bluer colours (greater
temperatures), this quantity tends to increase, especially for
solar metallicities.

\begin{table}
\centering
\begin{tabular}{c c c} \hline
$\Delta\feh$ & $\Delta T_{eff}$ for $\vs$ & $\Delta T_{eff}$ for
$\yv$ \\ \hline
$+0\longrightarrow-0.3$ & 39 & 45 \\
$-1\longrightarrow-1.3$ & 16 & 18 \\
$-2\longrightarrow-2.3$ & 8 & 9 \\ \hline
\end{tabular}
\caption{Mean variation $\Delta T_{eff}/\Delta\feh$ in K per 0.3
dex for $T_{eff}\sim5000$ K [$\vs\simeq\yv\simeq0.70$ for solar
metallicity] as a function of $\feh$.}
\label{gradients-feh-vilnius}
\end{table}

For both $\vs$ and $\yv$ the gradient $\Delta T_{eff}/\Delta\feh$
approaches zero as $\feh\longrightarrow-2$ which shows that our
formulae are reliable for halo stars even if their metallicities
have been roughly estimated. In addition, this effect is
independent of the colour values.

\subsection{Geneva colours}

A large quantity of stars have been observed with this photometric
system in just two observatories making the homogeneity of the
data excellent. We have found that approximately 60\% of the stars
with UBV colours have also Geneva photometry. Therefore, it will
be very useful to have calibrations for this system.

\begin{figure*}\centering
\includegraphics[bb=18 16 565 320,width=17cm]{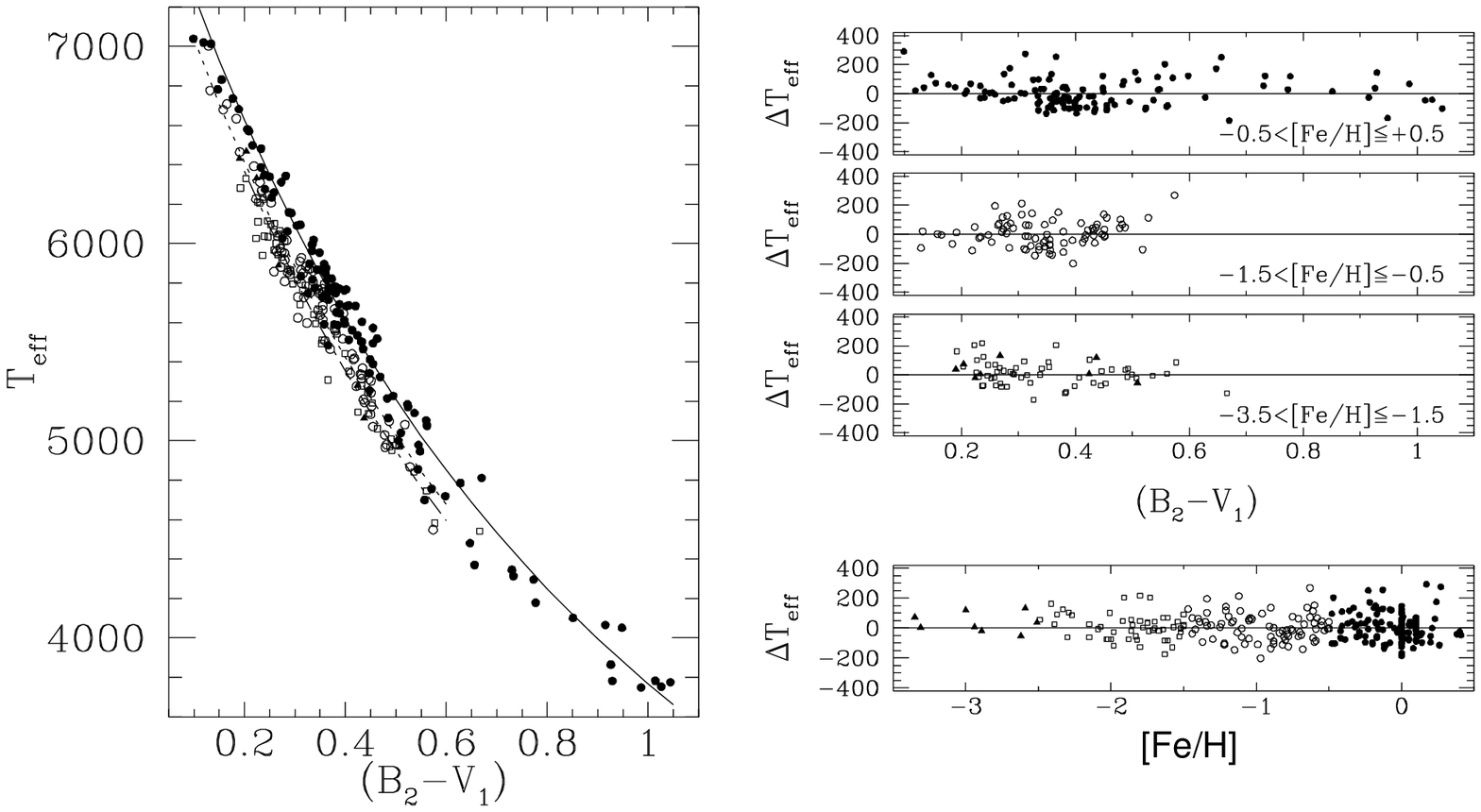}
\caption{The same as in Fig. \ref{fig:vilvs} for $T_{eff}$ versus
$\bbv$.} \label{fig:genb2v1}
\end{figure*}

For the $\theta_{eff}$-$\bbv$-$\feh$ relation we found the
following fit:
\begin{eqnarray}
\theta_{eff}=0.629+0.644\bbv+0.065\bbv^{2} \nonumber
\\ -0.033\bbv\feh-0.029\feh \nonumber \\
-0.010\feh^{2}\ ,\label{cali-b2v1}
\end{eqnarray}
obtained with data for 258 stars. The corresponding standard
deviation amounts to $\sigma(\theta_{eff})=0.016$ (90 K).

From Fig. \ref{fig:genb2v1} it is clear that Eq. (\ref{cali-b2v1})
is applicable only in the following intervals:
\[0.10<\bbv<1.05\ \ \ \textrm{for }\ -0.5<\feh<+0.5\ ,\]
\[0.10<\bbv<0.60\ \ \ \textrm{for }\ -1.5<\feh<-0.5\ ,\]
\[0.20<\bbv<0.60\ \ \ \textrm{for }\ -2.5<\feh<-1.5\ ,\]
\[0.20<\bbv<0.50\ \ \ \textrm{for }\ -3.5<\feh<-2.5\ .\]
It is also evident from this figure that the formula is
particularly reliable in the range $0.2<\bbv<0.6$ where the star
number density is high.

The stars that depart more than $2\sigma$ from the mean fit are:
G237-072, HR5901, HD3567, G251-054, G022-024, BD +71 31, HD111980,
G126-062, G055-044, HD101177, G076-068, G227-037 and HR6844.

The residuals of this fit as shown by Fig. \ref{fig:genb2v1}
suggest a slight tendency of Eq. (\ref{cali-b2v1}) to overestimate
the effective temperatures of the hottest solar-metallicity stars
and those of the $\feh\sim-3$ group.

With the same stars used to obtain Eq. (\ref{cali-b2v1}) we found
for the $\bbg$ colour index the following formula:
\begin{eqnarray}
\theta_{eff}=0.838+0.501\bbg-0.030\bbg^{2} \nonumber
\\ -0.044\feh-0.010\feh^{2}\ ,\label{cali-b2g}
\end{eqnarray}
applicable in the ranges:
\[-0.30<\bbg<1.10\ \ \ \textrm{for }\ -0.5<\feh<+0.5\ ,\]
\[-0.25<\bbg<0.40\ \ \ \textrm{for }\ -1.5<\feh<-0.5\ ,\]
\[-0.20<\bbg<0.40\ \ \ \textrm{for }\ -2.5<\feh<-1.5\ ,\]
\[-0.20<\bbg<0.30\ \ \ \textrm{for }\ -3.5<\feh<-2.5\ .\]
The standard deviation $\sigma(\theta_{eff})=0.014$ (86 K).

The following stars depart more than $2\sigma$ from the mean fit:
G021-022, HD3567, HD111980, HR7386, G076-068, G251-054, HD181007,
HD101177, G227-037 and HR6844.

\begin{figure*}\centering
\includegraphics[bb=18 16 565 320,width=17cm]{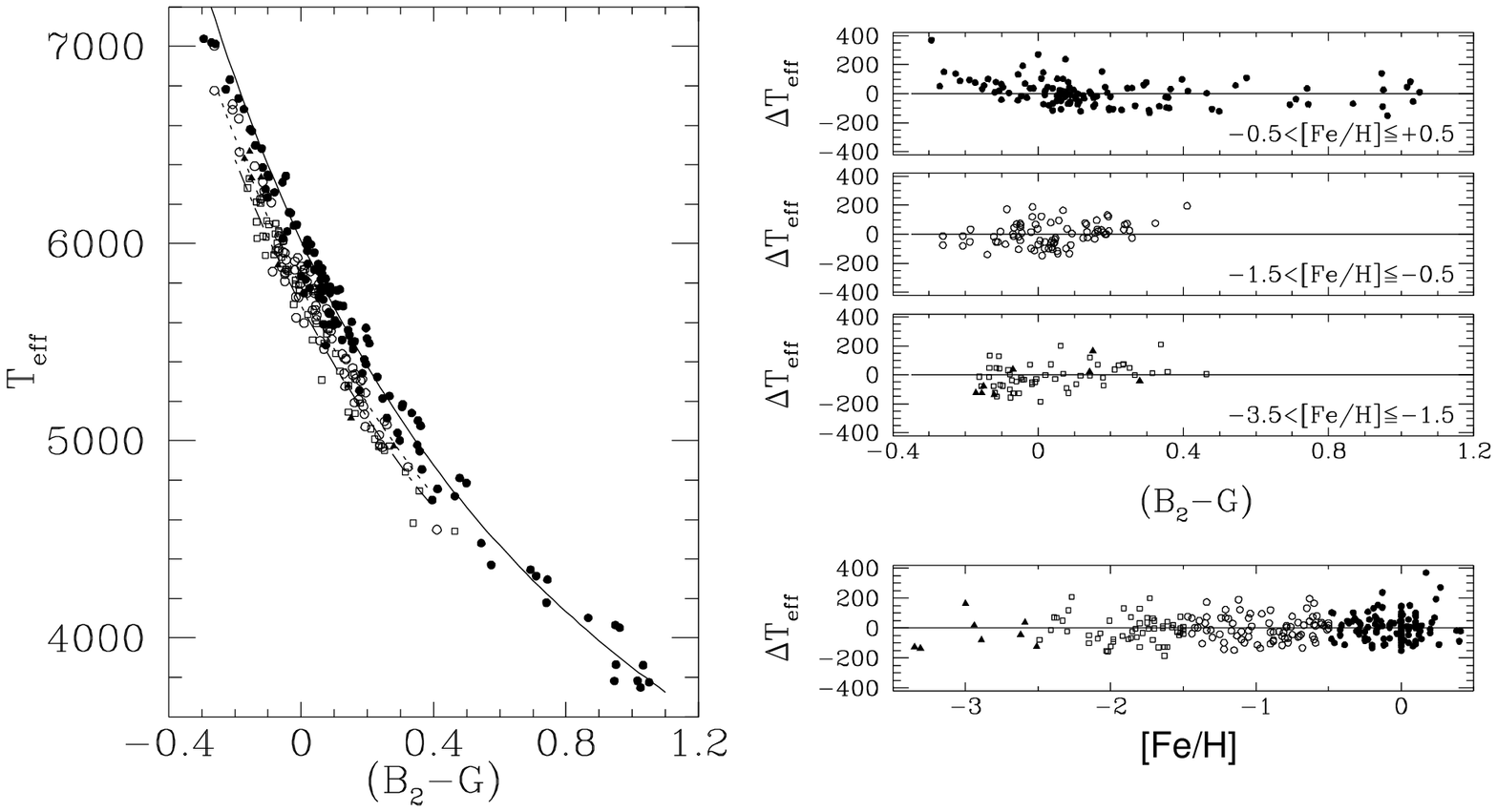}
\caption{The same as in Fig. \ref{fig:vilvs} for $T_{eff}$ versus
$\bbg$.} \label{fig:genb2g}
\end{figure*}

The sample and residuals of this fit are both shown in Fig.
\ref{fig:genb2g}. Here we find that Eq. (\ref{cali-b2g}) gives
higher temperatures (by $\sim100$ K) for solar-metallicity stars
with $T_{eff}>6400$~K.

There is only a slight dependence on $\feh$ for the $\Delta
T_{eff}/\Delta\bbg$ values, which vary from 50 K per 0.01 mag at
$\bbg\sim-0.2$ to 15 K per 0.01 mag at $\bbg\sim0.8$. In contrast
with this behavior, the gradient $\Delta T_{eff}/\Delta\feh$ is
higher for the hottest stars (it could be as high as 130 K per 0.3
dex at $\bbg\sim-0.2$ for solar metallicities) and lower for the
coolest ones (it is only about 50 K per 0.3 dex for $\bbg\sim0.8$
and $\feh\sim0$). Almost for any value of $\bbg$, the gradient
$\Delta T_{eff}/\Delta\feh$ approaches zero as
$\feh\longrightarrow-2$.

Following Strai\v{z}ys (\cite[p. 372]{straizys}) we have also
performed a calibration for the parameter
$\ttt\equiv\bbg-0.39\bbb$, which is almost independent of $\feh$
for F, G and early K stars. Data for 245 stars belonging to this
last group allowed us to obtain the following fit:
\begin{eqnarray}
\theta_{eff}=0.768+0.578\ttt+0.426\ttt^{2}-0.020\feh \nonumber
\\ -0.007\feh^{2}\ ,\label{cali-t}
\end{eqnarray}
with $\sigma(\theta_{eff})=0.012$ (75 K). This equation is
restricted to the ranges:
\[-0.10<\ttt<0.45\ \ \ \textrm{for }\ -0.5<\feh<+0.5\ ,\]
\[-0.05<\ttt<0.40\ \ \ \textrm{for }\ -1.5<\feh<-0.5\ ,\]
\[+0.00<\ttt<0.45\ \ \ \textrm{for }\ -2.5<\feh<-1.5\ ,\]
\[+0.00<\ttt<0.30\ \ \ \textrm{for }\ -3.5<\feh<-2.5\ .\]

\begin{figure*}\centering
\includegraphics[bb=18 16 565 320,width=17cm]{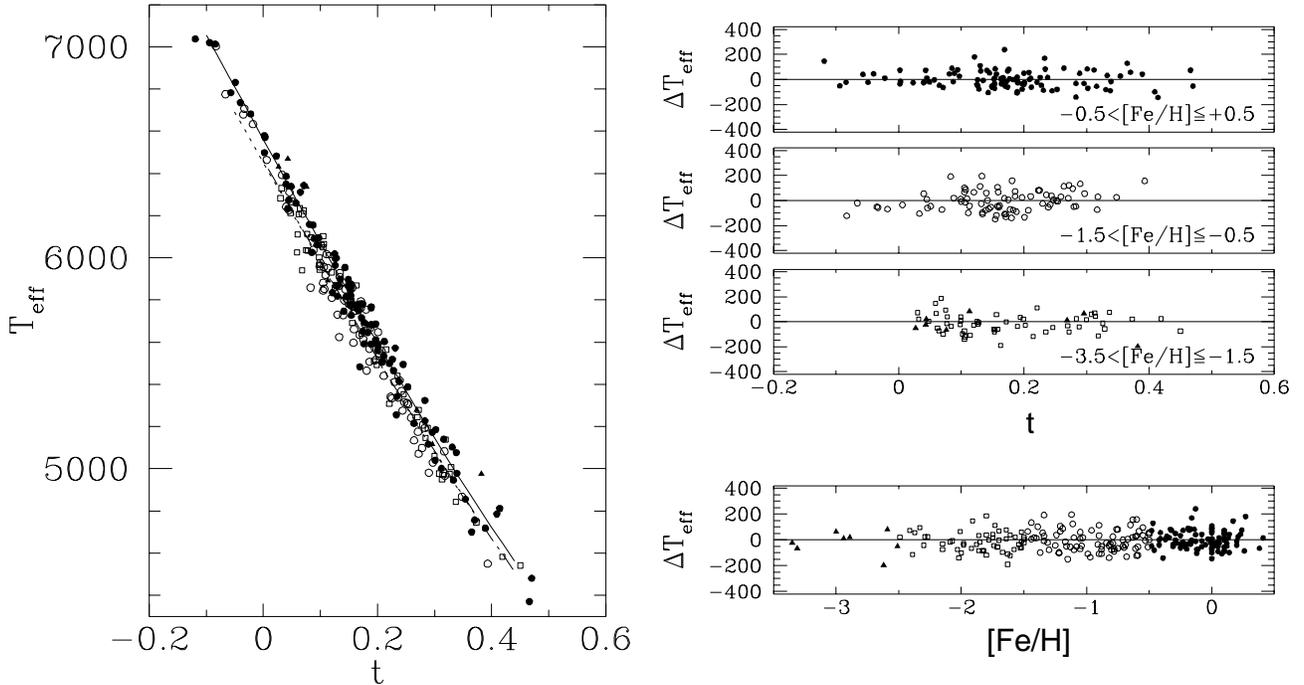}
\caption{The same as in Fig. \ref{fig:vilvs} for $T_{eff}$ versus
t.} \label{fig:gent}
\end{figure*}

The following stars deviate more than $2\sigma$ from the mean fit:
G014-024, G021-022, G076-068, G048-039, HR159, G227-037, G126-062,
HD3567, HD111980 and HD101177.

Unlike formulae (\ref{cali-b2v1}) and (\ref{cali-b2g}), the
present calibration does not produce systematic effects on any
metallicity or colour intervals as shown in Fig. \ref{fig:gent}.
From this figure it is also clear that the range $0<\ttt<0.4$
provides good estimates of $T_{eff}$ almost for any $\feh$ value.

The calibration formula (\ref{cali-t}) introduces a shift of about
50 K for an error of 0.01 mag in $\ttt$ if $-0.1<\ttt<0.2$ for any
value of the metallicity. This effect decreases monotonically as
cooler stars are considered. On the other hand, the gradient
$\Delta T_{eff}/\Delta\ttt$ vanishes for $\feh\sim-1.5$
independently of the $\ttt$ value and increases its absolute value
up to 80 K per 0.01 mag for the hottest stars going towards both
solar and very low metallicities.

\subsection{Cousins colours}

Since spectra of late type stars are dominated by emission at long
wavelengths, it is relatively easy to observe them with infrared
filters. In addition, infrared colours are good temperature
indicators because there are only slight blanketing effects in
this region of the spectrum.

\begin{figure*}\centering
\includegraphics[bb=18 16 565 320,width=17cm]{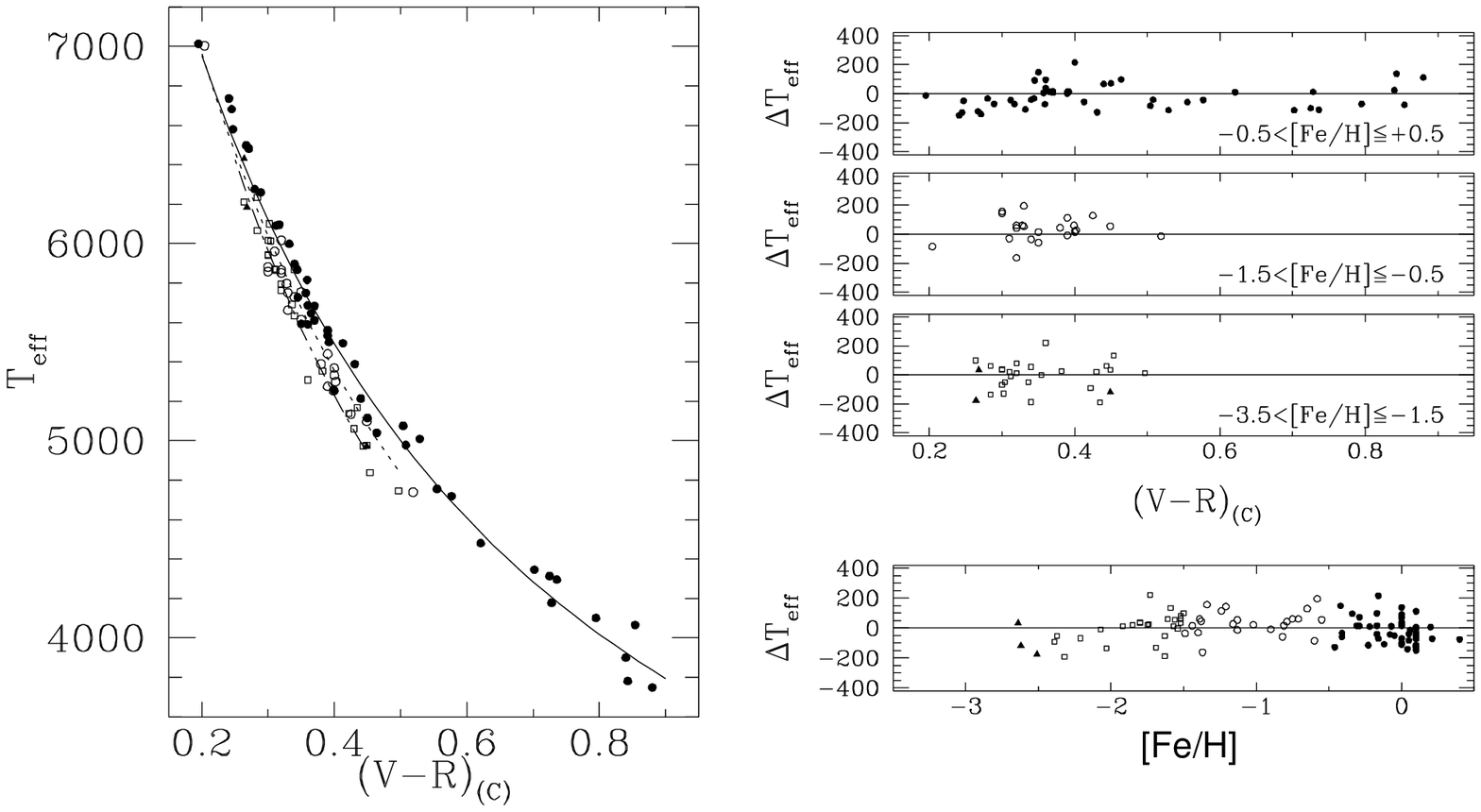}
\caption{The same as in Fig. \ref{fig:vilvs} for $T_{eff}$ versus
$\vr$.} \label{fig:ricvr}
\end{figure*}

\begin{figure*}\centering
\includegraphics[bb=18 16 565 320,width=17cm]{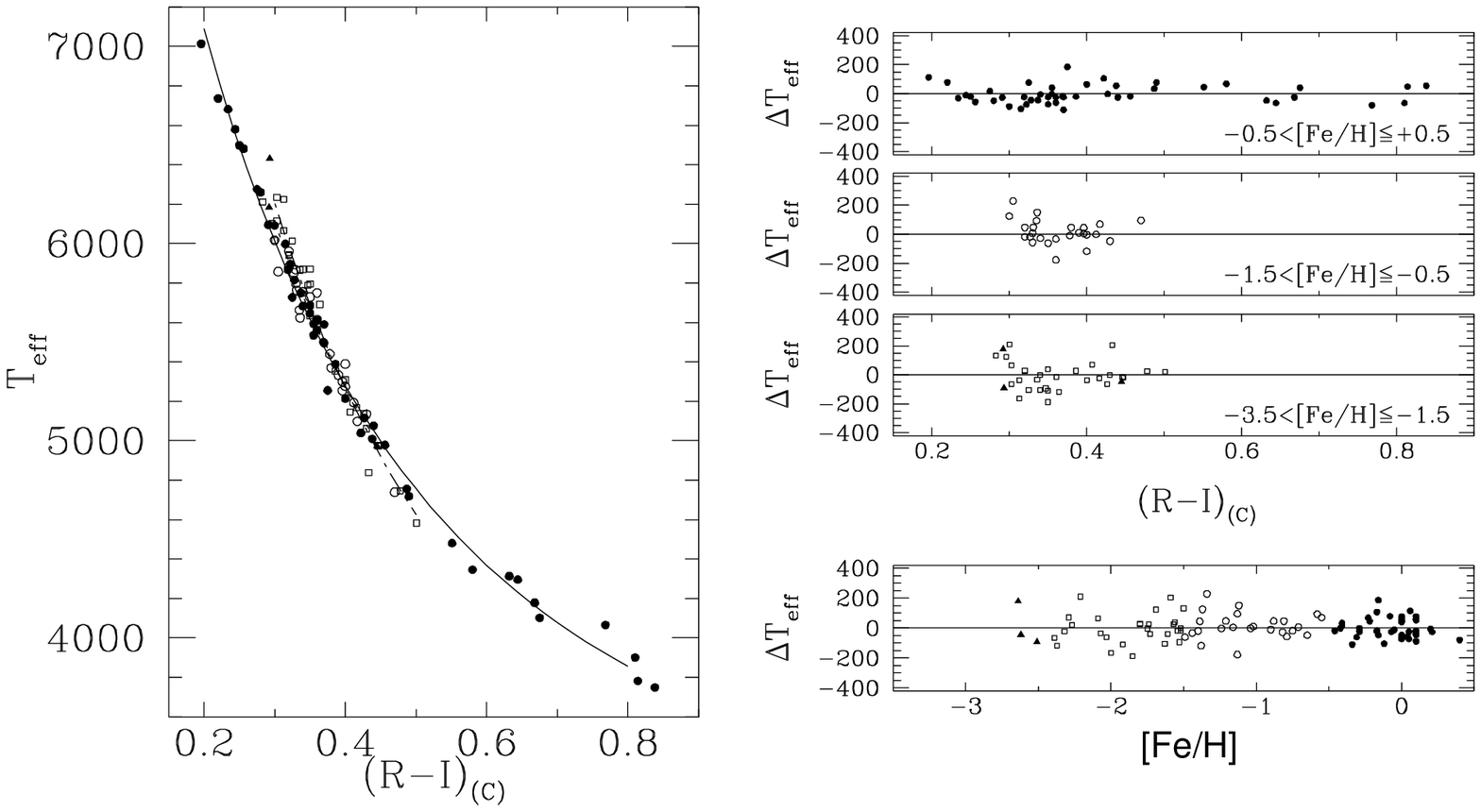}
\caption{The same as in Fig. \ref{fig:vilvs} for $T_{eff}$ versus
$\ri$.} \label{fig:ricri}
\end{figure*}

\begin{figure*}\centering
\includegraphics[bb=18 16 565 320,width=17cm]{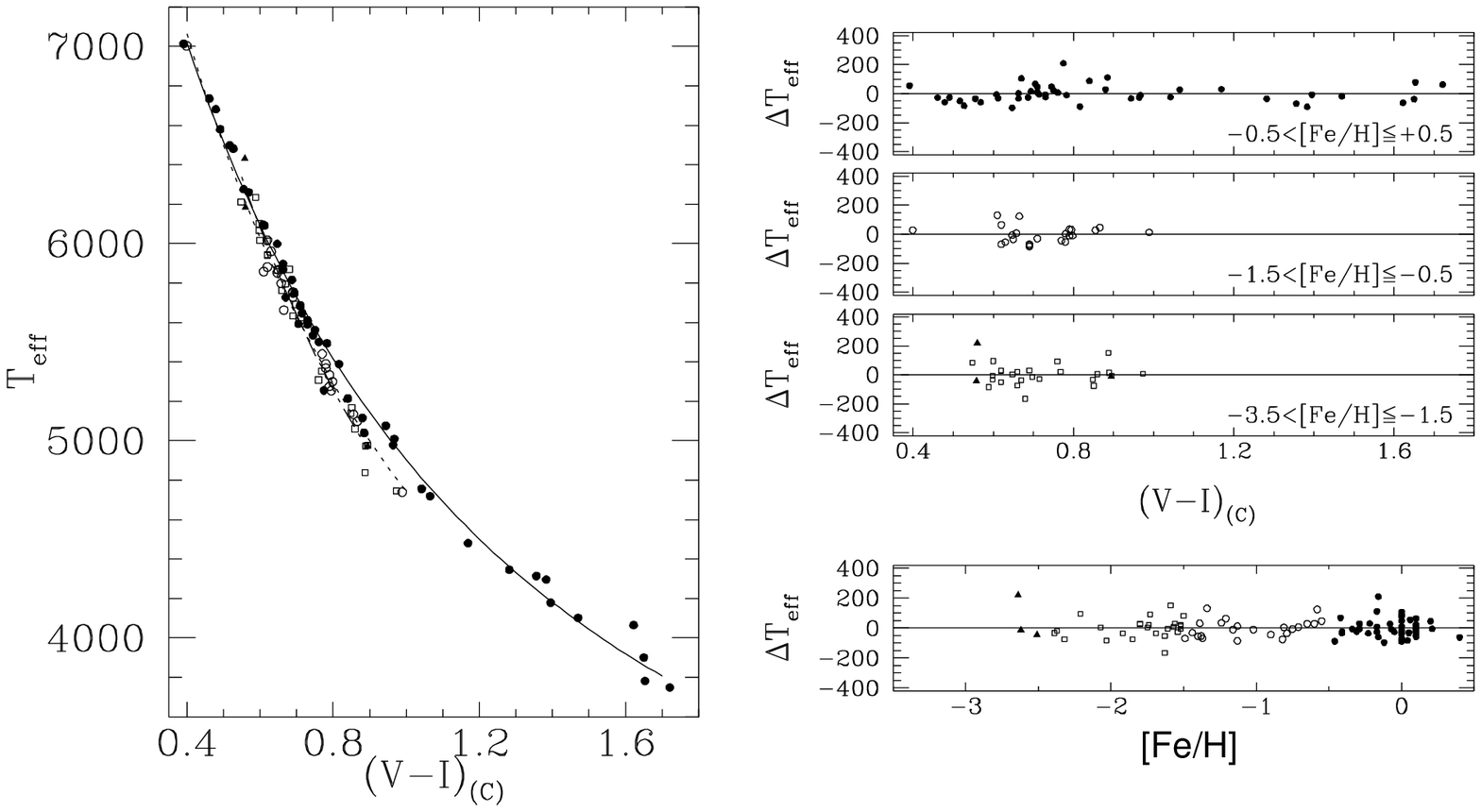}
\caption{The same as in Fig. \ref{fig:vilvs} for $T_{eff}$ versus
$\vi$.} \label{fig:ricvi}
\end{figure*}

For the $\vr$-$\theta_{eff}$-$\feh$ relation we found:
\begin{eqnarray}
\theta_{eff}=0.516+1.085\vr-0.203\vr^{2} \nonumber
\\ -0.120\vr\feh+0.025\feh \ ,\label{cali-vr}
\end{eqnarray}
using data for 97 stars. The standard deviation
$\sigma(\theta_{eff})=0.016$, which is equivalent to 91 K.

Eq. (\ref{cali-vr}) is applicable in the ranges
\[0.20<\vr<0.90\ \ \ \textrm{for }\ -0.5<\feh<+0.5\ ,\]
\[0.20<\vr<0.50\ \ \ \textrm{for }\ -1.5<\feh<-0.5\ ,\]
\[0.25<\vr<0.45\ \ \ \textrm{for }\ -2.5<\feh<-1.5\ ;\]
which are the colour index and metallicity intervals covered by
the sample, as shown in Fig. \ref{fig:ricvr}. Note that we have
not considered the $-3.5<\feh\leq-2.5$ interval because the only
three stars with $\feh<-2.5$ actually have $\feh\sim-2.6$.

The stars whose IRFM $T_{eff}$ depart more than $2\sigma$ from the
mean fit are: G180-058, G021-022, HD157089, HR159 and G251-054.

Errors in the metallicity estimative are not so important when
using Eq. (\ref{cali-vr}) if we are dealing with F stars since
$\Delta T_{eff}/\Delta\feh<20$ K per 0.3 dex. The gradient itself
is independent of $\feh$ for any value of the colour index and
reaches a maximum of 50 K per 0.3 dex for K dwarfs.

On the other hand, the observational errors on $\vr$ do affect the
temperature estimate for F stars since the variations $\Delta
T_{eff}/\Delta\vr$ can be as high as 90 K per 0.01 mag for
$\vr\sim0.22$. For the coolest stars this effect can be reduced to
30 K per 0.01 mag. The metallicity dependence of this quantity is
only slight.

We have 104 stars with measured $\ri$ colours which satisfy the
following formula:
\begin{eqnarray}
\theta_{eff}=0.422+1.557\ri-0.563\ri^{2} \nonumber
\\ -0.140\ri\feh+0.055\feh \ ,\label{cali-ri}
\end{eqnarray}
and cover the application ranges:
\[0.20<\ri<0.80\ \ \ \textrm{for }\ -0.5<\feh<+0.5\ ,\]
\[0.30<\ri<0.45\ \ \ \textrm{for }\ -1.5<\feh<-0.5\ ,\]
\[0.30<\ri<0.50\ \ \ \textrm{for }\ -2.5<\feh<-1.5\ .\]
The corresponding standard deviation $\sigma(\theta_{eff})=0.013$
which amounts to 80 K.

The stars whose temperatures obtained from this fit differ more by
than $2\sigma$ with their IRFM temperatures are: G089-014,
HD193901, HD74000, G026-012, HR159, G105-050, G088-027 and HD3567.

The relation between $T_{eff}$ and $\ri$ is not affected by the
star metallicity in the range $0.35<\ri<0.45$ where the variation
$\Delta T_{eff}/\Delta\feh\sim10$ K per 0.3 dex, which means an
error lower than 0.25\% in the estimated $T_{eff}$. For $\ri<0.35$
and $\ri>0.45$ this gradient increases to 30 K per 0.3 dex.

Except for the coolest stars, the gradient $\Delta
T_{eff}/\Delta\ri$ depends strongly on $\feh$, increasing its
absolute value from 90 K per 0.01 mag for $\feh\sim0$ to 120 K per
0.01 mag for $\feh\sim-2$ at $\ri\sim0.3$ for instance. For
$\ri\sim0.4$ an observational error of 0.01 mag would produce an
error of 70 K when using Eq. (\ref{cali-ri}) which means 1.3\%,
whilst for $\ri\sim0.7$ this reduces to 30 K (0.75\%).

The best temperature indicator for the Cousins system is the $\vi$
colour. Data for 97 stars satisfy the relation:
\begin{eqnarray}
\theta_{eff}=0.483+0.617\vi-0.072\vi^{2} \nonumber
\\ -0.066\vi\feh+0.023\feh \nonumber \\
-0.008\feh^{2} \ ,\label{cali-vi}
\end{eqnarray}
in the ranges:
\[0.40<\vi<1.70\ \ \ \textrm{for }\ -0.5<\feh<+0.5\ ,\]
\[0.40<\vi<1.00\ \ \ \textrm{for }\ -1.5<\feh<-0.5\ ,\]
\[0.55<\vi<0.90\ \ \ \textrm{for }\ -2.5<\feh<-1.5\ .\]

The standard deviation corresponding to Eq. (\ref{cali-vi}) is
$\sigma(\theta_{eff})=0.011$, which means the lowest dispersion
obtained in this work (64 K).

The following stars depart more than $2\sigma$ from the mean fit:
G021-022, HD3567, G105-050, HR159, G026-012.

The gradient $\Delta T_{eff}/\Delta\feh$ depends strongly on
colour and metallicity. Just to give an example, for $\vi\sim0.7$,
where the star number density is high (Fig. \ref{fig:ricvi}),
$\Delta T_{eff}/\Delta\feh$ vanishes for $\feh\sim-1.5$ but can be
as high as 60 K per 0.3 dex for solar metallicities. If we go to
$\vi\sim0.4$ the gradient now vanishes for solar metallicities and
is higher than 100 K per 0.3 dex for $\feh\sim-2$.

The ratio $\Delta T_{eff}/\Delta\vi$ monotonically decreases in
absolute value from 60 K per 0.01 mag to 15 K per 0.01 mag
depending on the mean value of $\vi$. For $\vi\sim0.7$ this
amounts to 35 K per 0.01 mag.

We should mention that from Figs. \ref{fig:ricvr}, \ref{fig:ricri}
and \ref{fig:ricvi} there is a slight tendency towards lower
temperatures if they are obtained from the corresponding
calibration formulae for the hottest solar-metallicity stars.
Despite this fact, we have found that our calibrations for the
Cousins colours do not introduce considerable systematic errors.

\subsection{DDO colours} \label{ddocolours}

\begin{figure*}\centering
\includegraphics[bb=18 16 565 320,width=17cm]{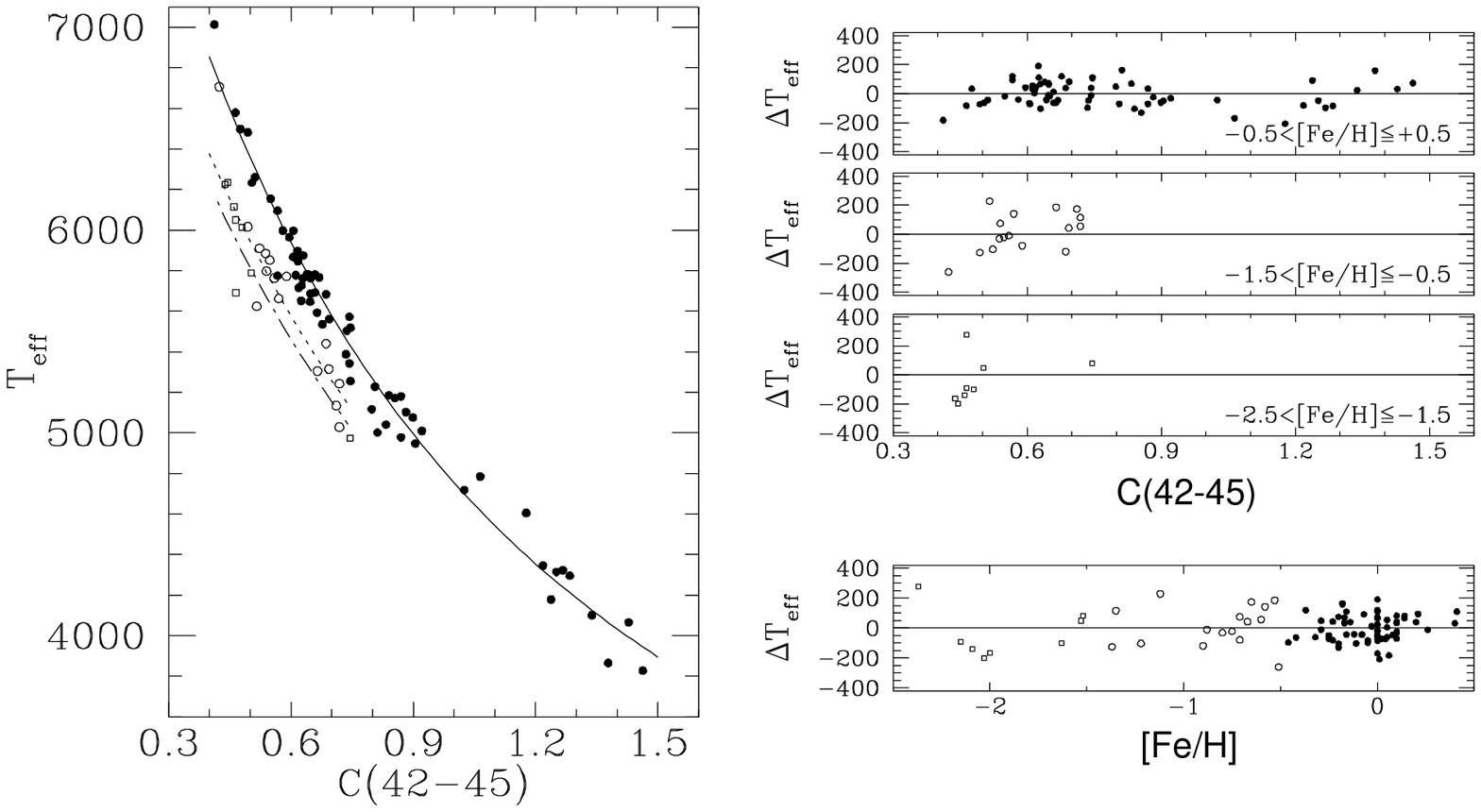}
\caption{The same as in Fig. \ref{fig:vilvs} for $T_{eff}$ versus
$\ddoa$.} \label{fig:4245}
\end{figure*}

Although data in this system were available for only 89 stars and
even though DDO colours are very sensitive to the star
metallicity, the need for an empirical DDO temperature scale
encouraged us to perform the following calibrations.

For the $\ddoa$ colour we found the following fit:
\begin{eqnarray}
\theta_{eff}=0.492+0.635[\ddoa]-0.067[\ddoa]^{2} \nonumber
\\ -0.073\feh-0.018\feh^{2} \ ,\label{cali-4245}
\end{eqnarray}
applicable in the ranges:
\[0.40<\ddoa<1.50\ \ \ \textrm{for }\ -0.5<\feh<+0.5\ ,\]
\[0.40<\ddoa<0.74\ \ \ \textrm{for }\ -1.5<\feh<-0.5\ ,\]
\[0.42<\ddoa<0.76\ \ \ \textrm{for }\ -2.5<\feh<-1.5\ .\]
Its standard deviation $\sigma(\theta_{eff})=0.018$ (103 K).

Stars whose IRFM temperatures depart more than $2\sigma$ from the
fit are: HR5447, HR5568, HD111980 and HD140283.

Finally, for the $\ddob$ colour, we obtained:
\begin{eqnarray}
\theta_{eff}=0.189+0.413[\ddob]-0.060\feh \nonumber
\\ -0.015\feh^{2} \ ,\label{cali-4248}
\end{eqnarray}
whose application ranges are:
\[1.30<\ddob<2.70\ \ \ \textrm{for }\ -0.5<\feh<+0.5\ ,\]
\[1.32<\ddob<1.82\ \ \ \textrm{for }\ -1.5<\feh<-0.5\ ,\]
\[1.32<\ddob<1.82\ \ \ \textrm{for }\ -2.5<\feh<-1.5\ .\]

The standard deviation of Eq. (\ref{cali-4248}) amounts to
$\sigma(\theta_{eff})=0.015$ (83 K) and only three stars depart
more than $2\sigma$ from the fit: HR5568, HD111980 and HD140283.

The general behavior of the gradients $\Delta T_{eff}/\Delta\ddoa$
and $\Delta T_{eff}/\Delta\ddob$ is very similar. Their absolute
values decrease for lower values of $\feh$ and $T_{eff}$, and they
both reach a minimum between 10 and 20 K per 0.01 mag for late K
stars ($\ddoa>1.2$, $\ddob>2.2$). For the hottest stars these
quantities depend slightly on metallicity, and are about 40 K and
30 K per 0.01 mag for $\ddoa$ and $\ddob$, respectively.

\begin{figure*}\centering
\includegraphics[bb=18 16 565 320,width=17cm]{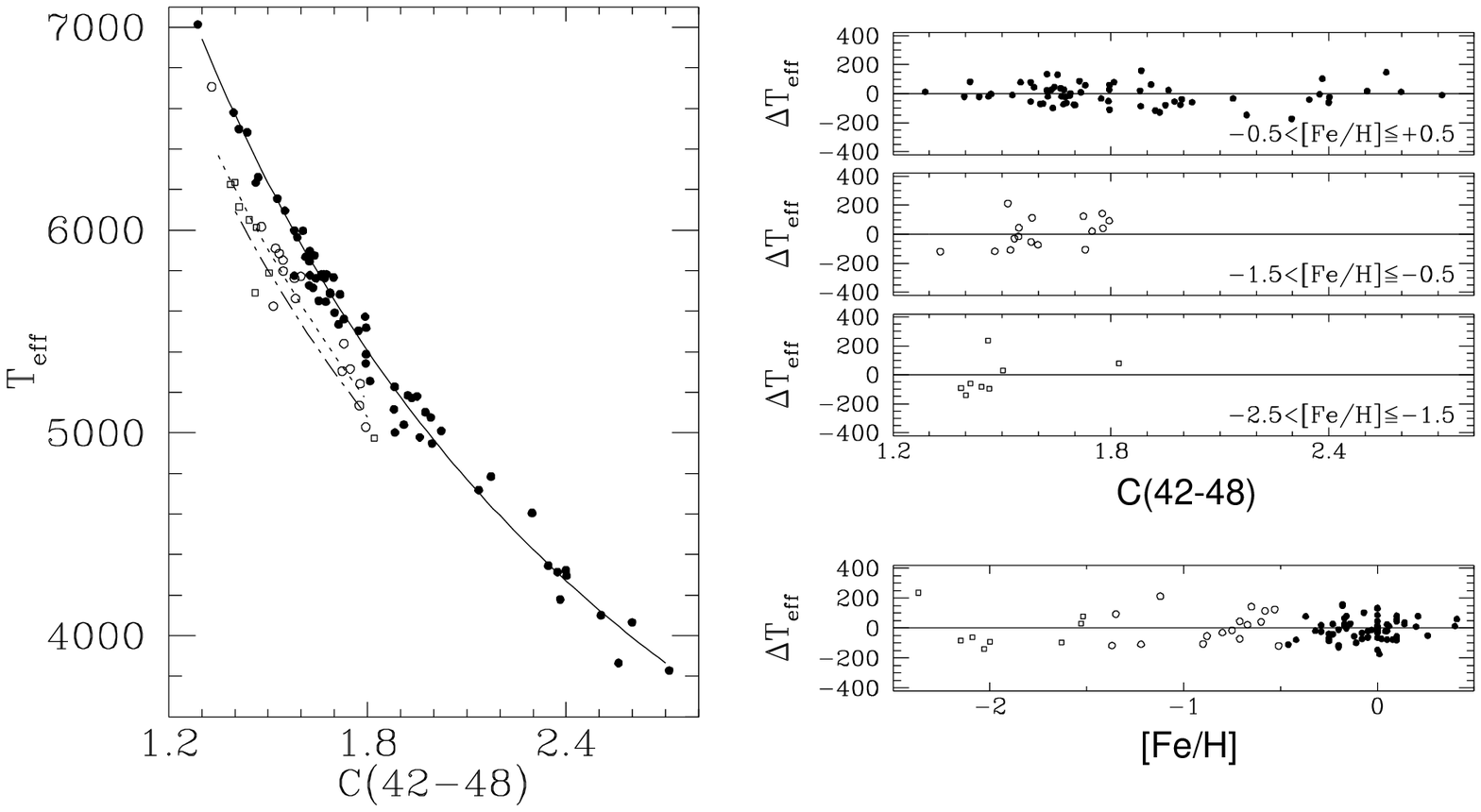}
\caption{The same as in Fig. \ref{fig:vilvs} for $T_{eff}$ versus
$\ddob$.} \label{fig:4248}
\end{figure*}

The high values found for the mean variations $\Delta
T_{eff}/\Delta\feh$ corresponding to Eqs. (\ref{cali-4245}) and
(\ref{cali-4248}) confirm the strong metallicity dependence of DDO
colours, which is also evident from the values of the coefficients
for $\feh$ and $\feh^{2}$ in the calibration formulae. For
$T_{eff}\sim6200$ K and solar metallicities, for example, $\Delta
T_{eff}/\Delta\feh\sim200$ K and 160 K per 0.3 dex for $\ddoa$ and
$\ddob$, respectively. For $\feh\sim-1$ these variations reduce to
aproximately 70 K. For lower temperatures $\Delta
T_{eff}/\Delta\feh$ decreases monotonically so that for
$T_{eff}\sim5300$ K and solar metallicity it is about 140 K and
100 K per 0.3 dex for $\ddoa$ and $\ddob$, respectively.

\subsection{Comparison with other calibrations} \label{comparison}

In Figs. \ref{fig:vgcomp} to \ref{fig:ddocomp} the present
calibrations for $\feh=0$ and $\feh=-1$ (solid lines) are compared
to previously published theoretical and empirical calibrations as
described below.

\subsubsection{Vilnius system}

\begin{figure*}\centering
\includegraphics[bb=18 16 565 320,width=17cm]{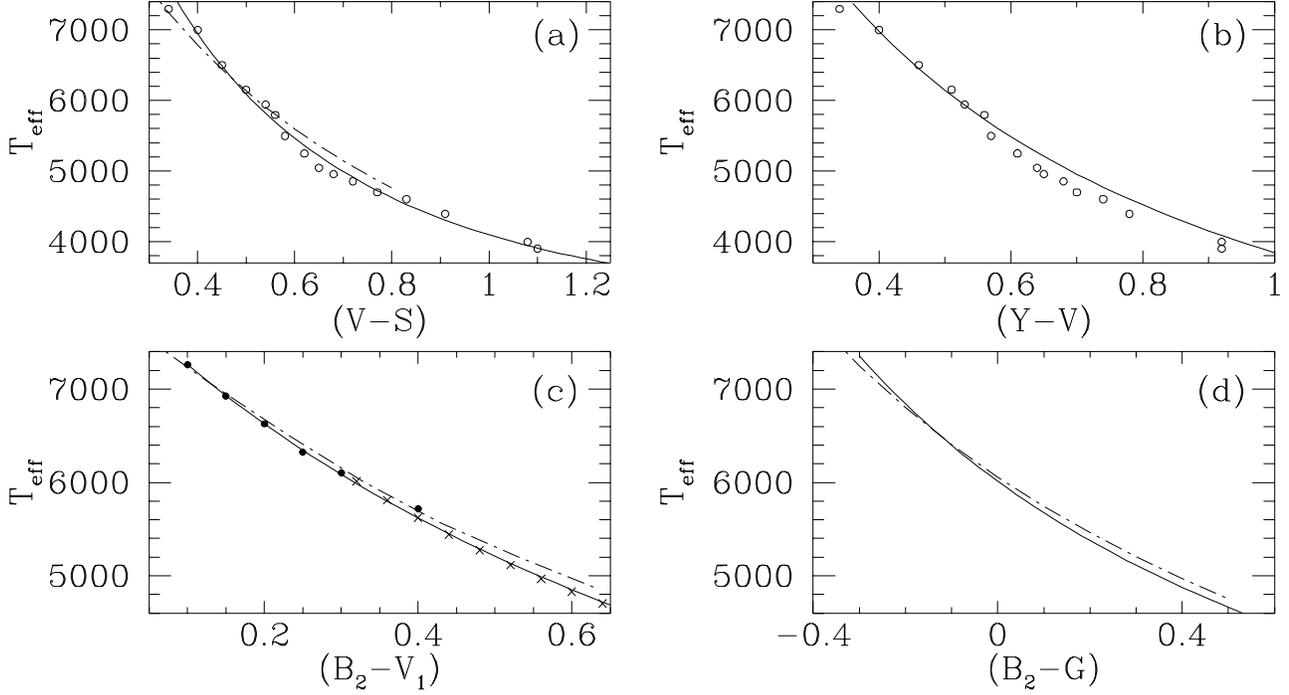}
\caption{\textbf{a-b.}Comparison of our calibrations (solid lines)
with those of Hauck \& K\"{u}nzli (\cite{hauck}) (dashed line) and
Strai\v{z}ys (\cite[p. 438]{straizys}) (open circles) for
$\feh=0$; \textbf{c-d.} Comparison of Eqs. (\ref{cali-b2v1}) and
(\ref{cali-b2g}) for $\feh=0$ (solid lines) to other calibrations:
Kobi \& North (\cite{kobi}) (filled circles), Hauck \& K\"{u}nzli
(\cite{hauck}) (dashed line) and Blackwell \& Lynas-Gray
(\cite{blackwell}) (crosses).} \label{fig:vgcomp}
\end{figure*}

The empirical work of Hauck \& K\"{u}nzli (\cite{hauck}; hereafter
HK96) for the relation between $T_{eff}$ and $\vs$ (shown in Fig.
\ref{fig:vgcomp}.a as a dashed line) is based on effective
temperatures obtained by several authors using direct and
semi-direct methods (including the IRFM). They fitted a linear
function of $\vs$ on $\theta_{eff}$ with 66 stars, some of them
out of the range covered in this work (namely, stars hotter than
7000 K). Since HK96 did not consider metallicity effects on their
calibration formula, we have assumed $\feh=0$ for the comparison.

For the hottest stars ($\vs\sim0.40$) the difference in derived
effective temperatures in our work and in HK96 is higher than 150
K, for the coolest ones ($\vs\sim0.80$) it is about $-140$ K
(these differences amount to 2.2\% and 2.9 \% respectively) whilst
it vanishes for $\vs\sim0.50$ so that the slopes are quite
different. This can be attributed to the calibration formulae used
in each work; the observed curvatures in our fits, for example,
are due mainly to the (colour)$^2$ term.

In Fig. \ref{fig:vgcomp}.a-b we also show the intrinsic colour
indices determined for luminosity class V stars of various
spectral types by averaging observed indices of unreddened and
dereddened stars of the same MK spectral type (Strai\v{z}ys
\cite[p. 438]{straizys}). The calibration of $T_{eff}$ in terms of
spectral type is also provided in Strai\v{z}ys (\cite{straizys}).
The agreement is good in the range 5800 K$<T_{eff}<7000$ K for
both $\vs$ and $\yv$ while at cooler temperatures, the maximum
difference in $\vs$ is about 150 K (2.9\%). For the calibration in
$\yv$ there is a shift of aproximately $-250$ K for the
Strai\v{z}ys (\cite{straizys}) work with respect to ours in the
range $0.56<\yv<0.92$ but both slopes are rather similar.

\subsubsection{Geneva system}

The calibration given by Kobi \& North (\cite{kobi}) is based on
synthetic Geneva colours obtained from older versions of Kurucz
models. It is compared to our calibration for the colour index
$\bbv$ in Fig. \ref{fig:vgcomp}.c. As usual, they corrected the
synthetic colours using standard stars so that they could be able
to reproduce the observed colours, especially for cool stars. This
correction was calculated only for $\feh=0$ but they assumed that
it was also valid for lower metallicities. Nevertheless, their
published results for $\feh<0$ are ambiguous and have not been
considered here.

Some of the standard stars mentioned above were used by HK96 to
calibrate this colour index and also the $\bbg$, hence the better
agreement between Kobi \& North (\cite{kobi}) and HK96 works for
the coolest stars. In total, HK96 used more than 140 stars for the
empirical calibrations shown in Fig. \ref{fig:vgcomp}.c-d. The
tendency of both calibrations with respect to ours are similar.
They give higher temperatures than ours for the coolest stars
whilst around $T_{eff}=6800$ K the difference vanishes. The
maximum difference amounts to 150 K for the $\bbv$ colour index
and 120 K for $\bbg$. In both cases this maximum is achieved at
$T_{eff}\sim4800$ K and is equivalent to 3.0 and 2.5 \%,
respectively.

In a recent work, Blackwell \& Lynas-Gray (\cite{blackwell}) used
the IRFM temperatures of a large sample of ISO (Infrared Space
Observatory) flux calibration stars and their Geneva colours
$\bbv$ corrected by interstellar extinction using Hipparcos
parallaxes and an average interstellar extinction law coupled with
extinction maps to calibrate the relation $T_{eff}=f\bbv$ where
$f$ is a second order polynomial. Their results are shown in Fig.
\ref{fig:vgcomp}.c as crosses. The agreement between our work and
theirs is excellent. We should also mention that the effect of
metallicity is considered in Blackwell \& Lynas-Gray
(\cite{blackwell}) but apparently valid only for $\feh>-0.6$.

\begin{figure*}
\centering
\includegraphics[bb=18 16 565 320,width=17cm]{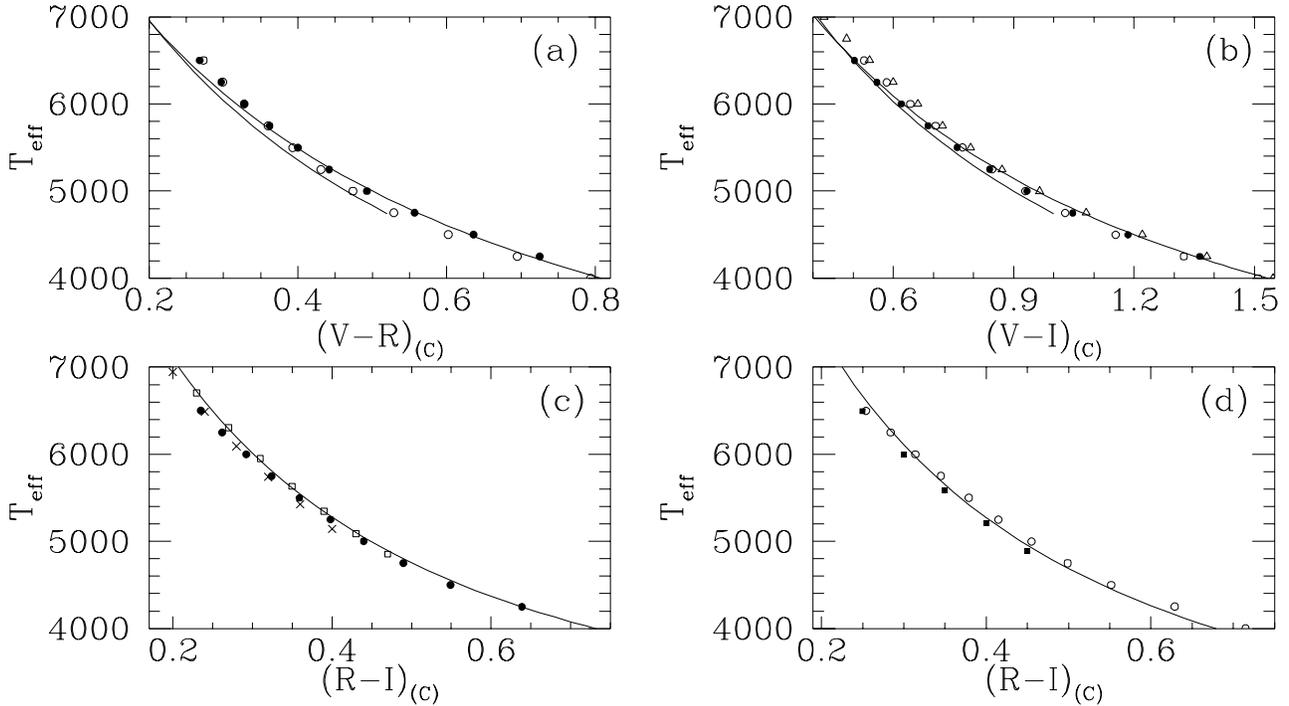}
\caption{\textbf{a-d.} Comparison of our calibrations for
$\feh=0,-1$ (solid lines) with those of Houdashelt et al.
(\cite{houdash}) for $\feh=0$ (filled circles) and $\feh=-1$ (open
circles), Bessell et al. (\cite{bessell}) for $\feh=0$
(triangles), Hauck \& K\"{u}nzli (\cite{hauck}) for $\feh=0$ (open
squares) and Carney et al. (\cite{carney}) for $\feh=0$ (crosses)
and $\feh=-1$ (filled squares).} \label{fig:ricomp}
\end{figure*}

\subsubsection{RI$_{\textrm{\tiny (C)}}$ system}

Houdashelt et al. (\cite{houdash}) have obtained
colour-temperature relations with improved MARCS stellar
atmosphere models (Gustafsson et al. \cite{gusta}; Bell et al.
\cite{bell}) after putting synthetic colours onto the
observational systems by means of a sample of field stars with
IRFM temperatures determined and a model for Vega, whose colours
fix the zero point corrections. These relations are shown in Figs.
\ref{fig:ricomp}.a-d as filled circles for $\feh=0$ and open
circles for $\feh=-1$. It is worth mentioning that the colour
calibrations were derived for $\feh=0$ stars so that their results
for lower metallicities are less reliable. Their calibrations for
solar metallicities seem to agree with ours for $T_{eff}<6000$~K
and for the whole range of temperatures in the case of $\vi$. For
the hottest stars the $\vr$ colours derived by Houdashelt et al.
(\cite{houdash}) are too red whilst the $\ri$ colours are too
blue. The maximum difference with our work is about 150 K.

In Fig. \ref{fig:ricomp}.b we show as open triangles the relation
between temperature and synthetic colours computed from spectra
obtained from overshoot (for $T_{eff}>6000$ K) and no-overshoot
(for $T_{eff}<6000$ K) models as given by Bessell et al.
(\cite{bessell}). The selection of overshoot and no-overshoot
models for the different ranges of temperatures alowed us to
improve the agreement with our work although it showed to be very
small. For effective temperatures below 5000 K our empirical
calibration is very close to this one but for the hottest stars
Bessell et al. (\cite{bessell}) provide $\vi$ colours that are too
red (the difference is about 0.03 mag). The agreement is also
good, though it is not shown, for the $\vi$ colours of the late K
dwarfs calculated on the basis of NMARCS models.

We also show in Fig. \ref{fig:ricomp}.c the empirical calibration
of HK96 (open squares), which, though being based on the effective
temperatures of only 26 solar-metallicity stars, agrees perfectly
well with the results of this work.

In their survey of proper motion stars, Carney et al.
(\cite{carney}) derived the metallicity dependent relation between
$T_{eff}$ and $\ri$ that is shown in Fig. \ref{fig:ricomp}.c for
$\feh=0$ (crosses) and Fig. \ref{fig:ricomp}.d for $\feh=-1$
(filled squares). Their temperature scale was based upon ``a
comparison of spectrophotometric scans of the Paschen continuum
with surface flux distributions computed with the same model
atmospheres used to calculate the synthetic spectra.'' In general,
their temperatures are lower than those calculated from our
formulae.

\begin{figure*}\centering
\includegraphics[bb=18 160 565 320,width=17cm]{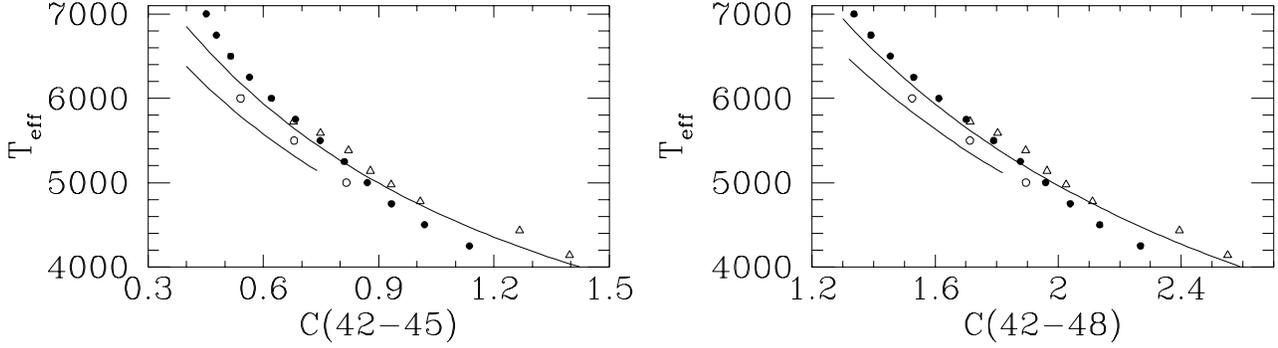}
\caption{Comparison of our calibrations for $\feh=0,-1$ (solid
lines) with those given by Tripicco \& Bell (\cite{tripicco}) for
$\feh=0$ (filled circles) and $\feh=-1$ (open circles) and
Clari\'{a} et al. (\cite{claria}) for $\feh=0$ (triangles).}
\label{fig:ddocomp}
\end{figure*}

\subsubsection{DDO system}

Finally, in Fig. \ref{fig:ddocomp} is shown the comparison between
our calibrations for the DDO colours and those given by Tripicco
\& Bell (\cite{tripicco}) and Clari\'a et al. (\cite{claria}). The
former of these (which is shown in Fig. \ref{fig:ddocomp} as
filled and open circles for $\feh=0$ and $\feh=-1$, respectively)
was based on MARCS models and used spectrophotometric scans for
several stars to put the synthetic colours onto the observational
system, thus allowing a model independent treatment. Tripicco \&
Bell (\cite{tripicco}) argue that the effect of metallicity on DDO
colours is so strong that ``DDO system is primarily a measure of
line, rather than continuum effects.'' This is something that can
be easily seen in Figs. \ref{fig:4245} and \ref{fig:4248} and was
already discussed in Sect. \ref{ddocolours}. For a given $T_{eff}$
in the range 5200-6200 K, the effect of metallicity is such that a
change from $\feh=0$ to $\feh=-1$ produces a decrease of 0.08 mag
in $\ddoa$ and 0.09 mag in $\ddob$ in the Tripicco \& Bell
(\cite{tripicco}) work. For the present calibration, the decrease
amounts to 0.1 mag in both cases. In this range there is a
reasonably good agreement but large systematic differences start
to appear as we go both to lower and higher temperature ranges. In
general, the slopes are not similar but the effect of $\feh$ on
DDO colours is well reproduced.

Spectroscopic observations of field and open cluster stars allowed
Clari\'a et al. (\cite{claria}) to calibrate the relation $\ddoc$
vs. $\ddoa$ for $\feh\sim0$. This colour-colour diagram was then
used to determine the mean values of DDO colours as a function of
spectral type so that the mean DDO effective temperature-MK
spectral type relation provided them the $T_{eff}$:DDO colours
calibration. The triangles in Fig. \ref{fig:ddocomp} show that our
temperatures are always lower than those derived by Clari\'a et al
(\cite{claria}), the maximum difference being about 200 K for K5
stars. There is a reasonable agreement around 5000 K.

\section{The empirical temperature scale}

\begin{figure}
\includegraphics[bb=15 166 573 700,width=8.5cm]{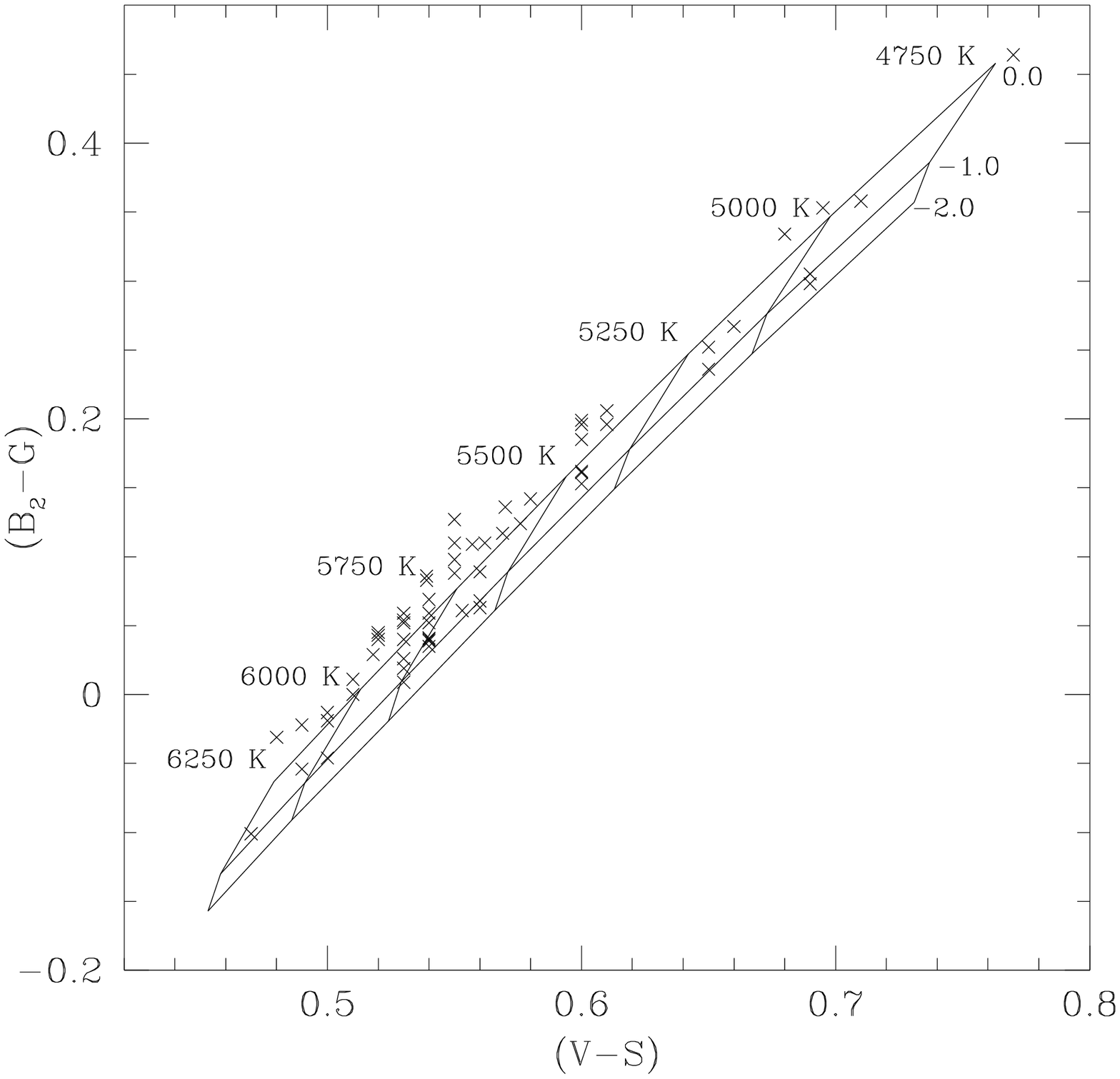}
\caption{$\vs$ versus $\bbg$, colour-colour diagram showing lines
of equal $T_{eff}$ and metallicity for $\feh=0,-1$ and $-2$.
Crosses correspond to observed colours for stars with
$-0.5<\feh<+0.5$.} \label{fig:vsb2g}
\end{figure}

\begin{figure}
\includegraphics[bb=15 166 573 700,width=8.5cm]{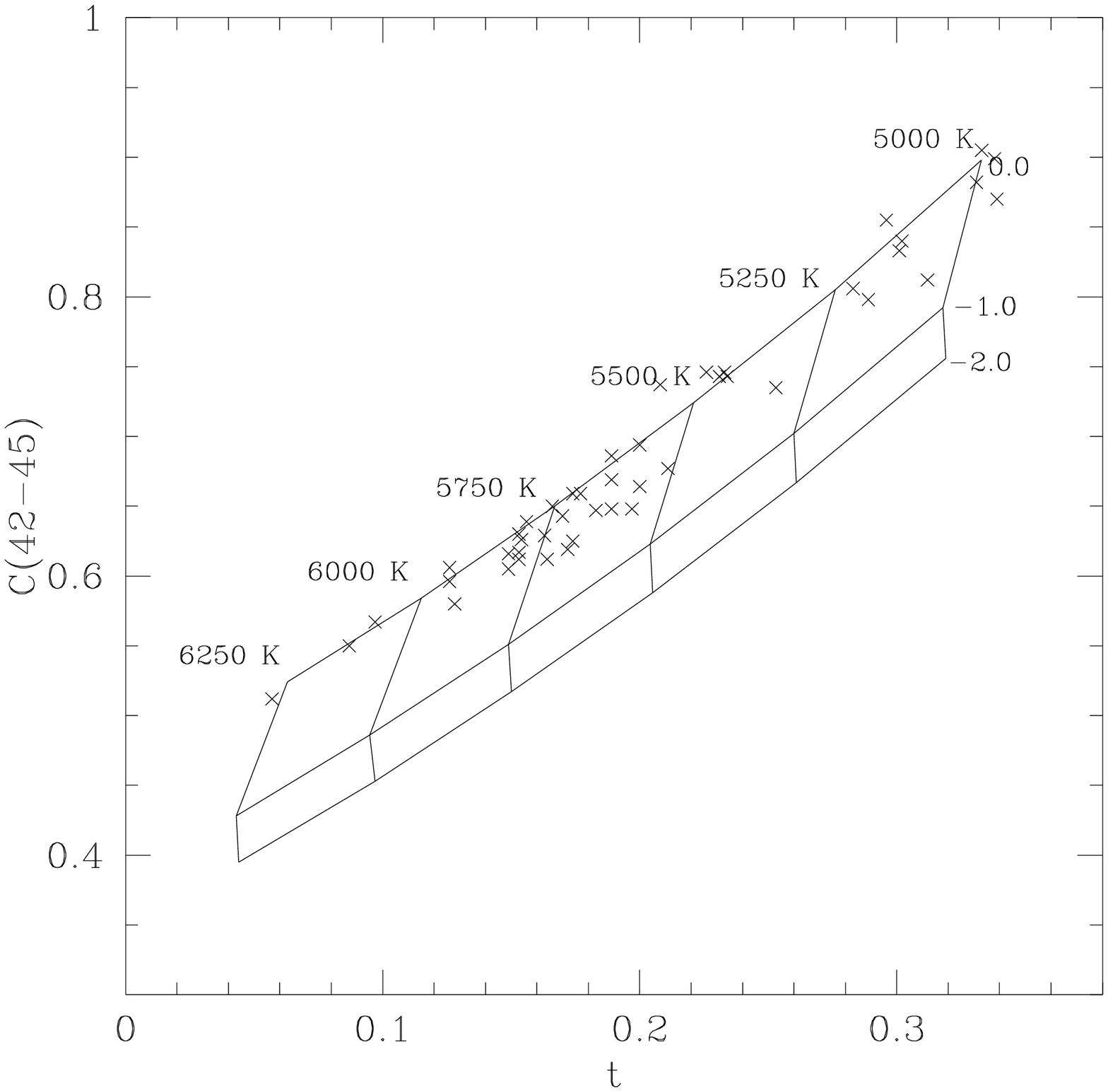}
\caption{The same as in Fig. \ref{fig:vsb2g} for $\ttt$ versus
$\ddoa$.} \label{fig:t4245}
\end{figure}

\begin{figure}
\includegraphics[bb=15 166 573 700,width=8.5cm]{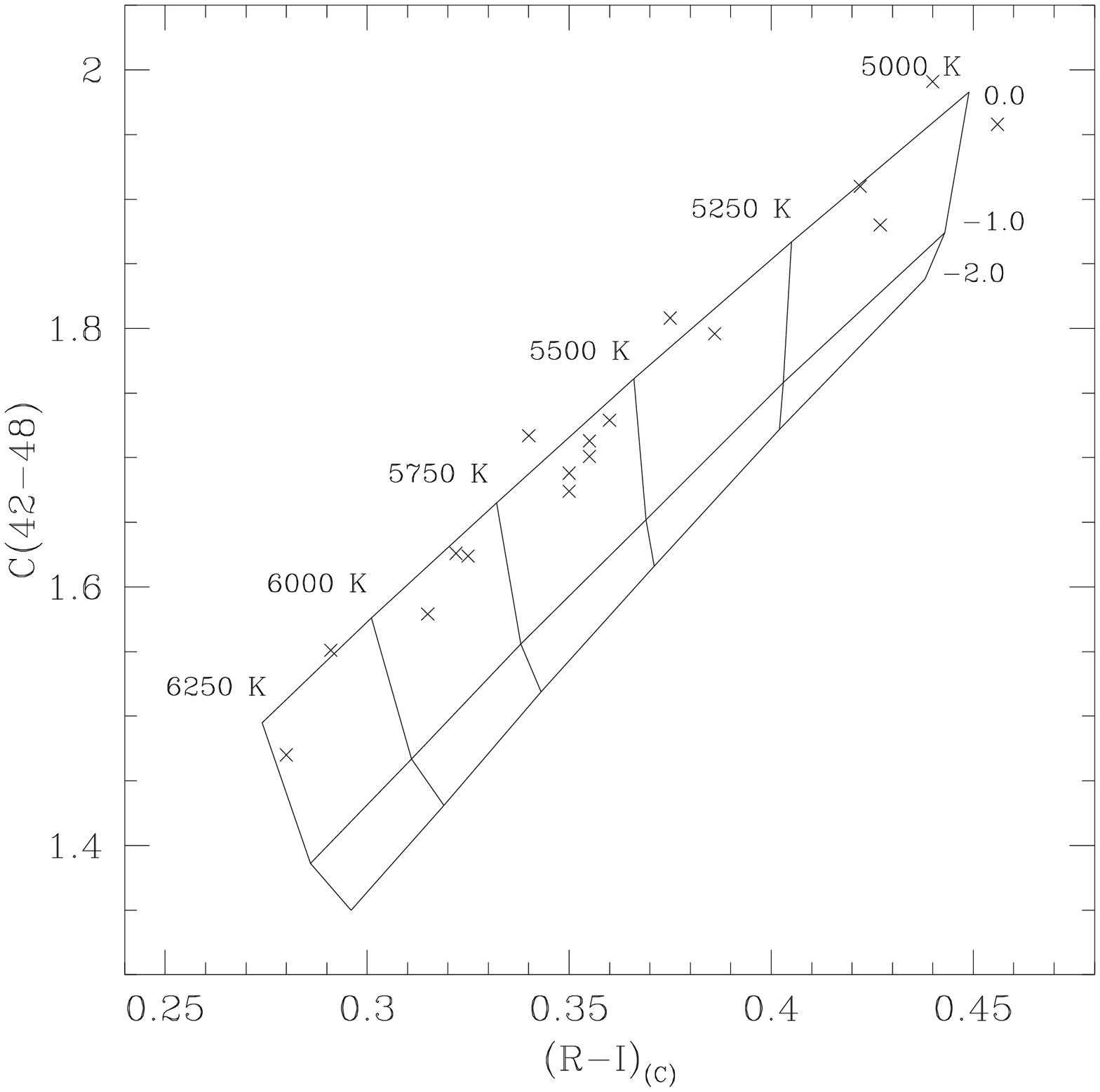}
\caption{The same as in Fig. \ref{fig:vsb2g} for $\ri$ versus
$\ddob$.} \label{fig:ri4248}
\end{figure}

\begin{figure}
\includegraphics[bb=15 166 573 700,width=8.5cm]{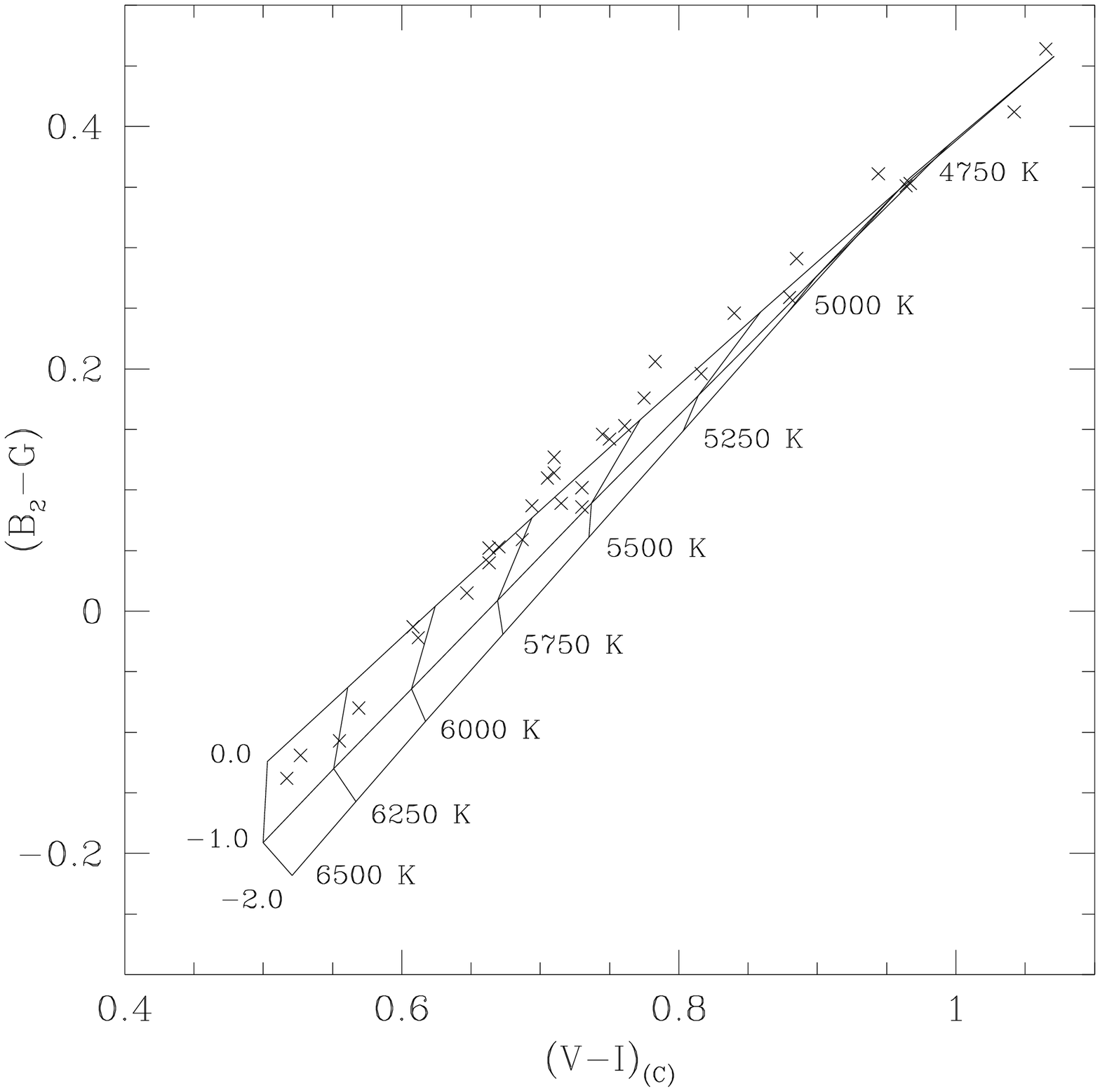}
\caption{The same as in Fig. \ref{fig:vsb2g} for $\vi$ versus
$\bbg$.} \label{fig:vib2g}
\end{figure}

Solving Eqs. (1) to (10) for the colour index as a function of
$T_{eff}$ and $\feh$ we obtained a set of intrinsic colours for
late type dwarfs. They are listed in Tables \ref{escala-vilnius}
to \ref{escala-ddo} for various metallicities. Because of their
usefulness, these results have been used to plot the colour-colour
diagrams shown in Figs. \ref{fig:vsb2g} to \ref{fig:vib2g}. Also
shown in these figures are the observed colours for stars with
$-0.5<\feh<+0.5$. The agreement is rather good since the observed
dispersion can be attributed primarily to the amplitude of the
metallicity range considered.

\begin{table*}
\centering
\begin{tabular}{c | r c c | r c c}\hline
 & & $\vs$ & & & $\yv$ & \\ \hline
$T_{eff}$ &  $\feh=0$ & $-1$ & $-2$ &  $\feh=0$ & $-1$ & $-2$
\\ \hline
3750    &   1.205   &   -   &   -   &   1.034   &   -   &   -   \\
4000    &   1.049   &   -   &   -   &   0.948   &   -   &   -   \\
4250    &   0.932   &   -   &   -   &   0.872   &   -   &   -   \\
4500    &   0.839   &   -   &   -   &   0.805   &   -   &   -   \\
4750    &   0.763   &   -   &   -   &   0.745   &   -   &   -   \\
5000    &   0.698   &   0.673   &   0.667   &   0.691   &   0.667 &   0.661   \\
5250    &   0.642   &   0.619   &   0.613   &   0.642   &   0.618 &   0.612   \\
5500    &   0.594   &   0.571   &   0.566   &   0.598   &   0.573 &   0.567   \\
5750    &   0.551   &   0.529   &   0.524   &   0.557   &   0.533 &   0.527   \\
6000    &   0.513   &   0.491   &   0.486   &   0.520   &   0.495 &   0.489   \\
6250    &   0.479   &   0.458   &   0.453   &   0.486   &   0.461 &   0.455   \\
6500    &   0.448   &   0.427   &   -   &   0.454   &   0.430   & -   \\
6750    &   0.420   &   0.400   &   -   &   0.425   &   -   &   - \\
\hline
\end{tabular}
\caption{Intrinsic colours in the Vilnius system as a function of
$\feh$.} \label{escala-vilnius}
\end{table*}

\begin{table*}
\centering
\begin{tabular}{c | r c c | r c c | r c c }\hline
 & & $\bbv$ & & & $\bbg$ & & & $\ttt$ & \\ \hline
$T_{eff}$ &  $\feh=0$ &  $-1$ & $-2$ & $\feh=0$ & $-1$ & $-2$ & $\feh=0$ & $-1$ & $-2$   \\ \hline
3750    &   1.008   &   -   & -   &   1.080   &   -   &   -   & -   &   -   &   -   \\
4000    &   0.898   &   -   &   -   &   0.890   &   -   &   -   & -   &   -   &   -   \\
4250    &   0.800   &   -   &   -   &   0.726   &   -   &   -   &  -   &   -   &   -   \\
4500    &   0.711   &   -   &   -   &   0.583   &   -   & - &   0.456   &   -   &   0.443   \\
4750    &   0.631   &   0.578   &   0.555   &   0.458   &   0.386 &   0.357   &   0.393   &   0.379   &   0.380   \\
5000    &   0.557   &   0.507   &   0.487   &   0.347   &   0.276  &   0.247   &   0.333   &   0.318   &   0.319   \\
5250    &   0.490   &   0.442   &   0.424   &   0.247   &   0.178  &   0.149   &   0.276   &   0.260   &   0.261   \\
5500    &   0.428   &   0.382   &   0.367   &   0.158   &   0.089  &   0.061   &   0.221   &   0.204   &   0.205   \\
5750    &   0.370   &   0.327   &   0.314   &   0.077   &   0.009  &   -0.019  &   0.167   &   0.149   &   0.150   \\
6000    &   0.317   &   0.276   &   0.265   &   0.004   &   -0.064 &   -0.091  &   0.115   &   0.095   &   0.097   \\
6250    &   0.268   &   0.229   &   0.220   &   -0.063  &   -0.130 &   -0.157  &   0.063   &   0.043   &   0.044   \\
6500    &   0.222   &   0.185   &   -   &   -0.124  &   -0.191  &  -   &   0.013   &   -0.010  &   -   \\
6750    &   0.179   &   0.144   &   -   &   -0.180  &   -0.247  &  -   &   -0.038  &   -0.062  &   -   \\
7000    &   0.139   &   0.105   &   -   &   -0.232  &   -   &   -  &   -0.089  &   -   &   -   \\ \hline
\end{tabular}
\caption{The same as in Table \ref{escala-vilnius} for the Geneva
system.}
\label{escala-geneva}
\end{table*}

\begin{table*}
\centering
\begin{tabular}{c | r c c | r c c | r c c }\hline
 & & $\vr$ & & & $\ri$ & & & $\vi$ & \\ \hline
$T_{eff}$ &  $\feh=0$ &  $-1$ & $-2$ & $\feh=0$ & $-1$ & $-2$ & $\feh=0$ & $-1$ & $-2$   \\ \hline
3750    &   0.922   &   -   &   -   &   -   &   -   &   -   &   - &   -   &   -   \\
4000    &   0.808   &   -   &   -   &   0.732   &   -   &   -   & 1.534   &   -   &   -   \\
4250    &   0.712   &   -   &   -   &   0.638   &   -   &   -   & 1.353   &   -   &   -   \\
4500    &   0.631   &   -   &   -   &   0.563   &   -   &   -   & 1.201   &   -   &   -   \\
4750    &   0.561   &   -   &   -  &   0.501   &   -   & -   &   1.071   &   -   &   -   \\
5000    &   0.500   &   0.466   &   0.439   &   0.449   &   0.443 &   0.438   &   0.958   &   0.899   &   0.879   \\
5250    &   0.447   &   0.419   &   0.397   &   0.405   &   0.403 &   0.402   &   0.859   &   0.814   &   0.803   \\
5500    &   0.399   &   0.377   &   0.360   &   0.366   &   0.369 &   0.371   &   0.772   &   0.737   &   0.735   \\
5750    &   0.356   &   0.339   &   0.326   &   0.332   &   0.338 &   0.343   &   0.694   &   0.669   &   0.673   \\
6000    &   0.317   &   0.305   &   0.296   &   0.301   &   0.311 &   0.319   &   0.624   &   0.607   &   0.617   \\
6250    &   0.283   &   0.274   &   0.268   &   0.274   &   -   & -   &   0.561   &   0.551   &   0.567   \\
6500    &   0.251   &   0.246   &   -   &   0.249   &   -   &   - &   0.503   &   0.500   &   -   \\
6750    &   0.222   &   0.220   &   -   &   0.227   &   -   &   - &   0.451   &   0.453   &   -   \\
7000    &   0.195   &   -   &   -   &   0.207   &   -   &   -   & 0.403   &   0.410   &   -   \\ \hline
\end{tabular}
\caption{The same as in Table \ref{escala-vilnius} for the
RI$_{(\textrm{\tiny C})}$ system.}
\label{escala-cousins}
\end{table*}

\begin{table*}
\centering
\begin{tabular}{c | r c c | r c c}\hline
 & & $\ddoa$ & & & $\ddob$ & \\ \hline
$T_{eff}$ &  $\feh=0$ & $-1$ & $-2$ &  $\feh=0$ & $-1$ & $-2$
\\ \hline
4000    &   1.423   &   -   &   -   &   2.593   &   -   &   -   \\
4250    &   1.260   &   -   &   -   &   2.414   &   -   &   -   \\
4500    &   1.122   &   -   &   -   &   2.254   &   -   &   -   \\
4750    &   1.002   &   -   &   -   &   2.112   &   -   &   -   \\
5000    &   0.898   &   0.792   &   0.756   &   1.983   &   1.874 &   1.838   \\
5250    &   0.805   &   0.702   &   0.667   &   1.867   &   1.758 &   1.722   \\
5500    &   0.724   &   0.623   &   0.588   &   1.761   &   1.652 &   1.616   \\
5750    &   0.650   &   0.551   &   0.517   &   1.665   &   1.556 &   1.519   \\
6000    &   0.584   &   0.486   &   0.453   &   1.576   &   1.467 &   1.431   \\
6250    &   0.524   &   0.428   &   0.395   &   1.495   &   1.386 &   1.350   \\
6500    &   0.470   &   -   &   -   &   1.420   &   -   &   -   \\
6750    &   0.420   &   -   &   -   &   1.350   &   -   &   -   \\
\hline
\end{tabular}
\caption{The same as in Table \ref{escala-vilnius} for the DDO
system.}
\label{escala-ddo}
\end{table*}

For the sun ($T_{eff}=5777$ K, $\feh=0$), our temperature scale
yields: $\vs_{\odot}=0.547$, $\yv_{\odot}=0.553$,
$\bbv_{\odot}=0.365$, $\bbg_{\odot}=0.069$, $\ttt_{\odot}=0.161$,
$\vr^{\odot}=0.352$, $\ri^{\odot}=0.328$, $\vi^{\odot}=0.686$,
$\ddoa_{\odot}=0.643$ and $\ddob_{\odot}=1.655$.  These results
are in good agreement with the solar colours compiled in HK96 and
almost with every calibration mentioned in Sect. \ref{comparison}
for it is clear from Figs. \ref{fig:vgcomp} to \ref{fig:ddocomp}
that the differences vanish for $T_{eff}\sim5800$ K. It is also
interesting to note that these solar colours are similar to the
colours of the ``solar twin'' 18 Sco (Porto de Mello \& da Silva
\cite{soltwin}): $\vs=0.54$, $\yv=0.57$, $\bbv=0.385$,
$\bbg=0.089$, $\ttt=0.17$, $\vr=0.353$, $\ri=0.335$, $\vi=0.686$,
$\ddoa=0.651$ and $\ddob=1.674$.

\section{Conclusions}

We have derived the relation  between $T_{eff}$, $\feh$ and
several colours for different photometric systems and have found
good agreement with previously published calibrations. We have
given detailed information about the application ranges and made
several warnings on the use of our calibration formulae. The
standard deviations amount from 64 K for $\vi$ to 126 K for $\yv$
and stars whose IRFM effective temperatures depart by more than
$2\sigma$ from the mean fit have always been listed.

The solar colours derived in this work have proved to be similar
to the colours of the solar twin 18 Sco so now these colours can
be used to look for new solar twins.

Along with AAM96b work, our calibrations provide an homogeneous
(IRFM) temperature scale for cool dwarfs which can be used to
explore the capability of models to reproduce the observations. A
similar extension for giants is also necessary.

\begin{acknowledgements}
This research has been supported by CONCYTEC (156-2002) and CSI
(Consejo Superior de Investigaciones - UNMSM).
\end{acknowledgements}

\end{document}